\documentclass[3p,times,authoryear]{elsarticle}

\usepackage{tabularx}
\usepackage{xltabular}
\usepackage{orcidlink}
\usepackage{enumitem}
\usepackage[usestackEOL]{stackengine}
\usepackage{amsmath,amssymb,amsfonts}
\usepackage{appendix}
\newcommand{\fref}[1]{Figure~\ref{#1}}
\newcommand{\tref}[1]{Table~\ref{#1}}

\newcommand{\cref}[1]{Chapter~\ref{#1}}
\newcommand{\sref}[1]{Section~\ref{#1}}

\usepackage{changes}

\journal{Journal of Informetrics}

\begin{document}

\begin{frontmatter}



\title{Integration vs segregation: network analysis of interdisciplinarity in funded and unfunded research on infectious diseases} 

\author[label1]{Anbang Du\corref{cor1}\orcidlink{0000-0003-4049-2778}}
\cortext[cor1]{Corresponding author}

\ead{ad1u21@soton.ac.uk}
\affiliation[label1]{organization={School of Electronic and Computer Science, University of Southampton},
            city={Southampton},
            postcode={SO17 1BJ},
            country={UK}}

\author[label2]{Michael Head\orcidlink{0000-0003-1189-0531}}
\ead{m.head@soton.ac.uk}
\affiliation[label2]{organization={Faculty of Medicine, University of Southampton},
            city={Southampton},
            postcode={SO17 1BJ},
            country={UK}}

\author[label1]{Markus Brede}
\ead{markus.brede@soton.ac.uk}



\begin{abstract}
Interdisciplinary research fuels innovation. In this paper, we examine the interdisciplinarity of research output driven by funding. Considering 36 major infectious diseases, we model interdisciplinarity through temporal correlation networks based on funded and unfunded research from 1995-2022. Using hierarchical clustering, we identify coherent periods of time or regimes characterised by important research topics like vaccinations or the Zika outbreak. We establish that funded research is less interdisciplinary than unfunded research, but the effect has decreased markedly over time. In terms of network growth, we find a tendency of funded research to focus on readily established connections leading to compartmentalisation and conservatism. In contrast, unfunded research tends to be exploratory and bridge distant knowledge leading to knowledge integration. Our results show that interdisciplinary research on prominent infectious diseases like HIV and tuberculosis tends to have strong bridging effects facilitating global knowledge integration in the network. At the periphery of the network, we observe the emergence of vaccination-related and Zika-related knowledge clusters, both with limited systemic impact. We further show that despite the surge in publications related to COVID-19, its systematic impact on the disease network remains relatively low. Overall, this research provides a generalisable framework to examine the impact of funding in interdisciplinary knowledge creation. It can assist in priority setting, for example with horizon scanning for new and emerging threats to health, such as pandemic planning. Policymakers, funding agencies, and research institutions should consider revamping evaluation systems to reward interdisciplinary work and implement mechanisms that promote and support intelligent risk-taking.
\end{abstract}


\begin{highlights}
\item We present a framework to analyse funded vs unfunded interdisciplinary research (IDR)
\item Funded IDR compartmentalises by reinforcing existing research clusters.
\item Unfunded IDR is more exploratory by linking distant knowledge.
\item As a result, funded IDR segregates while unfunded IDR integrates.
\item We explore how major events like SARS, Zika, and COVID-19 shape IDR.
\end{highlights}

\begin{keyword}
Interdisciplinary Research \sep Knowledge Integration \sep Research funding \sep Conservatism \sep Temporal Network Analysis


\end{keyword}

\end{frontmatter}


\section{Introduction}

Interdisciplinary research (IDR), a process of knowledge integration \citep{rafols_diversity_2010}, has been seen as a source of creativity and innovativeness \citep{Rousseau_knowledge_2019}. Various policy and funding initiatives have been developed to facilitate IDR \citep{wang_is_2015}. However, it has been discovered that highly interdisciplinary \citep{bromham_interdisciplinary_2016} and highly novel \citep{boudreau_looking_2016} ideas that integrate knowledge in unprecedented ways \citep{uzzi_atypical_2013,fontana2020new} tend to be penalised during grant proposal evaluation. Given such conservatism towards highly interdisciplinary ideas at the stage of funding application, does it carry further to research output generated by funding? It is an important question to ask as proposals represent the intention of a scientific idea while publications represent the realisation \citep{packalen_nih_2020}, i.e., when the intention of knowledge integration is discouraged, what happens to the realisation of knowledge integration? 

Funding's influences have been explored in a wide range of aspects, including on research methods and goals \citep{serrano_velarde_way_2018,luukkonen_negotiated_2016}, team composition \citep{davies_research_2022}, scientific productivity \citep{hottenrott_fishing_2017}, and scholarly \citep{roshani2021relationship,mosleh2022scientific,coccia2024general,coccia2024research}, technological and social impact \citep{heyard_value_2021,yang_unveiling_2024} of the associated research output. Funded research has been found to exhibit higher scholarly impact than its unfunded counterpart in various fields \citep{roshani2021relationship,mosleh2022scientific,coccia2024general,coccia2024research}.

Of those studies examining various aspects related to research output produced by funding \citep{JACOB20111168,wahls_national_2019,packalen_nih_2020,arora_impact_2006,heyard_value_2021,arora_reputation_2000,benavente_impact_2012,yang_unveiling_2024,roshani2021relationship,mosleh2022scientific,coccia2024general,coccia2024research}, limited attention has been paid to examine its interdisciplinarity (ID). In addition, studying funded research output beyond individual funder level at scale remains a gap \citep{yang_unveiling_2024}.

In this paper, we address the above gaps. By identifying all funded and unfunded publications in the field of infectious disease from 1995 to 2022 and considering a set of major infectious diseases to be the unit of analysis, we propose a novel temporal network approach to characterise and compare the ID of funded and unfunded research. We aim to answer the following research questions (RQ)s: 
\begin{enumerate}[label=RQ\arabic*:]
    \item Can we identify coherent periods of time, i.e., temporal regimes, in the evolution of ID in funded and unfunded research?
    \item What are the main trends in the time-dependence of ID in funded and unfunded research?
    \item Since IDR is discouraged at grant application, is funded research less interdisciplinary in general compared to unfunded research? What are the most under/over-funded interdisciplinary research areas?
    \item What roles does research into prominent infectious diseases like HIV and coronavirus play in terms of interdisciplinary knowledge generation? 
    \item What is the effect of important events like the 2002-2004 SARS outbreak or the COVID-19 pandemic on the ID of funded and unfunded infectious disease research? 
\end{enumerate}
The paper is structured as follows: \sref{Literature Review} provides a brief review of research on ID and funded research, \sref{Data and Methods} introduces the data source and computational implementation of IDR measures, \sref{Results} presents the main results and \sref{Discussion} summarises the results and discusses real-world implications, limitations, and future work.

\section{Literature Review}
\label{Literature Review}
We briefly review research on IDR in \sref{IDR: Overview} and the central concepts of quantitative IDR measures (diversity, coherence and intermediation) in \sref{Central Concepts}; then, we will review research on funded research and highlight the gaps in the field in \sref{Funded research and interdisciplinarity}.

\subsection{Interdisciplinary Research: Overview}
\label{IDR: Overview}
According to the National Academies of Sciences of the USA \cite{NAP11153}, IDR is "a mode of research by teams or individuals that integrates information, data, techniques, tools, perspectives, concepts and/or theories from two or more disciplines or bodies of specialized knowledge to advance fundamental understanding or to solve problems whose solutions are beyond the scope of a single discipline or area of research practice."
IDR, as a process of knowledge integration, generates creativity and innovativeness \citep{NAP11153,rafols_diversity_2010,Rousseau_knowledge_2019}. Current research on IDR has concluded that IDR could take place cognitively or socially \citep{glanzel_various_2022}, meaning the integration either happens in a scientist's mind \citep{rafols_diversity_2010} or in a social process \citep{abramo_identifying_2012,abramo_interdisciplinary_2017}, i.e., formation of a team of researchers with different expertise, experiences, or academic background. Based on these two perspectives, researchers have been exploring a broad range of questions on IDR: from the quantification of the ID of articles \citep{rafols_diversity_2010}, journals \citep{leydesdorff_betweenness_2007,rodriguez_disciplinarity_2017,leydesdorff_betweenness_2018}, scientists \citep{porterMeasuringResearcherInterdisciplinarity2007,leahey2017prominent}, institutes \citep{rafols_how_2012,soos_beyond_2012,biancani2018superstars}, grant proposals \citep{bromham_interdisciplinary_2016, nichols_topic_2014}, and research fields \citep{leydesdorffLocalEmergenceGlobal2011}, to mapping the global structure of science \citep{leydesdorff_global_2009,rafols_science_2010}, further to the scholarly and social impact of IDR \citep{okamura_interdisciplinarity_2019,hu_interdisciplinary_2024,shi_impact_2009,wang_interdisciplinarity_2015,yegros-yegros_does_2015,biancani2018superstars,leahey2017prominent}. For example, \cite{rafols_diversity_2010} used diversity and coherence as a framework to compare the ID of a set of bionanoscience articles. \cite{leydesdorff_global_2009} and \cite{leydesdorff_betweenness_2007} analysed the ID of journals based on betweenness centrality and diversity. \cite{porterMeasuringResearcherInterdisciplinarity2007} proposed the measures of integration, reach and specialisation and used them to measure the ID of $47$ researchers. \cite{soos_beyond_2012} analysed and compared the disciplinary structure of a sample of Hungarian Research Institutions based on the science overlay maps \citep{rafols_science_2010}. 

IDR may lack institutional appreciation due to mono-disciplinary academic structures \citep{Rousseau_knowledge_2019} and is poorly rewarded by funders \citep{bromham_interdisciplinary_2016} due to disciplinary-based evaluations \citep{Rousseau_knowledge_2019,woelert_paradox_2013,rylance2015global,fontana2022interdisciplinarity}. \cite{rafols_how_2012} compared the ID of Innovation Studies (IS) units with leading Business \& Management Schools (BMS) in the UK. Despite IS units being more interdisciplinary, they were disadvantaged in evaluation and obtaining resources due to the bias of the Association of Business Schools’ (ABS) journal rankings favouring mono-disciplinary research. \cite{biancani2018superstars} found interdisciplinary research centres have superior performance in knowledge production, instruction, collaboration and funding acquisition. \cite{leahey2017prominent} found that being interdisciplinary is a high risk high reward endeavour in a scientific career, i.e., a trade-off between productivity and scientific impact. IDR has been found to exhibit higher scholarly impact \citep{okamura_interdisciplinarity_2019} but \cite{shi_impact_2009}, \cite{wang_interdisciplinarity_2015} and \cite{yegros-yegros_does_2015} reported mixed results across different measures \citep{wang_interdisciplinarity_2015,yegros-yegros_does_2015} and different fields \citep{shi_impact_2009}. IDR has also been found to attract more attention from policy documents \citep{hu_interdisciplinary_2024}.

\subsection{Central Concepts in Quantitative Research on IDR}
\label{Central Concepts}
In IDR studies considerable effort has been used to develop quantitative measures to inform policymakers, research managers, evaluators and sociologists of science \citep{wagner_approaches_2011}. Here we briefly revisit relevant concepts in the quantification of IDR, but see \cite{wagner_approaches_2011} and \cite{wang_consistency_2020} for more comprehensive reviews.

There are three central concepts that quantitative IDR measures have been based on: diversity, coherence, and intermediation. Diversity refers to the difference in the bodies of knowledge that are integrated, and it consists of variety, balance, and disparity \citep{stirling_general_2007}. The Rao-Stirling (RS) index is one of the most widely-used diversity-based metrics \citep{wang_consistency_2020,nichols_topic_2014}, and \cite{zhang_diversity_2016} proposed the Hill-type measure as an improvement to the low discriminating power of the RS index. In addition, diversity measures adapted from evolutionary biology \citep{bromham_interdisciplinary_2016} have also been used to measure ID, e.g., \cite{bromham_interdisciplinary_2016} used the Phylogenetic Species Evenness (PSE), a measure of the biodiversity of species, to capture ID. 

Coherence describes the extent of relatedness of bodies of knowledge \citep{rafols_diversity_2010,rafols_knowledge_2014}. Coherence can be defined and used differently \citep{rafols_diversity_2010,soos_beyond_2012,rafols_how_2012}, and measures of coherence (and also intermediation) are still at an exploratory stage \citep{rafols_how_2012}. When coherence was first proposed \citep{rafols_diversity_2010}, it was used to measure the overall compactness of an article's knowledge structure. Rafols used a bibliographic coupling network to represent the knowledge structure. The nodes were the article's references and the links measured connection strengths (based on shared references of references). The coherence of each article is then computed as the mean linkage strength of the network. In contrast, \cite{soos_beyond_2012} focused on the coherence of research institutions, expressed by the sum of weighted distances of the Web of Science (WOS) Subject Category\footnote{Formerly known as the Institute for Scientific Information (ISI) Subject Category.} (SC) in an institute's publication profile. Distances between WOSSCs are based on the global map of science \citep{leydesdorff_global_2009}, and the weights assigned to the distances are the intensity of interactions between WOSSCs in an institute's publication profile, i.e., cooccurrences. \cite{soos_beyond_2012} further suggested multimodality could also reflect coherence, where multimodality is defined as the size distribution of connected components of the network of SCs and quantified by the Shannon-Wiener entropy \citep{shannon_mathematical_nodate}. \cite{rafols_knowledge_2014} later proposed tentatively that the concept of coherence of interconnected bodies of knowledge consists of density (number of links), intensity (strength of links) and disparity (degree of difference in two bodies of knowledge that links bridge), but its added value and feasibility remain uncertain \citep{Rousseau_knowledge_2019}. 

Intermediation refers to the ability to link distant bodies of knowledge \citep{leydesdorff_betweenness_2007}. Betweenness was first proposed by \cite{leydesdorff_betweenness_2007} as a measure of intermediation. Leydesdorff thinks journals with high betweenness are more interdisciplinary due to their capability to relate otherwise non-interacting journals. The average Local Clustering Coefficient (LCC) was also used to capture the effect of intermediation \citep{rafols_how_2012,soos_beyond_2012}. The LCC of a body of knowledge reflects to what extent its neighbouring fields are directly connected (or integrated), indicating the tendency of transitivity in the neighbouring field, i.e., to what extent intermediation “results” in integration \citep{soos_beyond_2012}. In \cite{rafols_how_2012}, this concept was operationalised slightly differently by assigning weights (the proportion of publications from each body of knowledge) to each LCC. \cite{soos_beyond_2012} further proposed the network diameter (or the maximal shortest path length) to capture both coherence and intermediation. A low diameter means nodes are quickly reachable from each other, i.e., highly coherent. On the other hand, a high diameter indicates some constituent fields are distant but are still linked through a ``mediator", implying the role of intermediation.

\subsection{Funded research and interdisciplinarity}
\label{Funded research and interdisciplinarity}
The influence of funding on research has been studied in various aspects (see \cite{thelwall_what_2023} for further detail), including research methods or goals \citep{serrano_velarde_way_2018,luukkonen_negotiated_2016}, team composition \citep{davies_research_2022}, researcher, team and institutional level productivity \citep{heyard_value_2021,hottenrott_fishing_2017}, output types \citep{thelwall_what_2023}, and scholarly \citep{roshani2021relationship,mosleh2022scientific,coccia2024general,coccia2024research}, technological and social impacts of outputs \citep{heyard_value_2021,yang_unveiling_2024}. 

Funding sources, especially those that routinely issue topic-focused calls \citep{viergever_10_2016}, seem to have the greatest influence on most science and health research which are resource intensive \citep{thelwall_what_2023,whitley_impact_2018}; however, the degree of such influence depends on individual funders' requirements \citep{serrano_velarde_way_2018,luukkonen_negotiated_2016,thelwall_what_2023}. 

Funding has demonstrated great value to knowledge creation by facilitating scientific productivity \citep{heyard_value_2021,hottenrott_fishing_2017}. Funding also plays a key role in knowledge diffusion \citep{roshani2021relationship,mosleh2022scientific,coccia2024general,coccia2024research}, where funded research tends to be more often cited than unfunded research in a wide variety of fields including computer science, economics, physics, chemistry and medicine. Moreover, funded research also tends to produce higher technological impact (through citation by patent) and social engagement (through citation by Tweets) \citep{yang_unveiling_2024}.

Studies on funded research output have been mostly based on publications associated with individual funders, e.g., the National Institutes of Health (NIH) \citep{JACOB20111168,wahls_national_2019,packalen_nih_2020,yang_unveiling_2024}, the National Science Foundation (NSF) \citep{arora_impact_2006,yang_unveiling_2024}, the Swiss National Science Foundation \citep{heyard_value_2021}, the Italian National Research Council \citep{arora_reputation_2000}, the Chilean National Science and Technology Research Fund \citep{benavente_impact_2012}. The recent development of the SciSciNet \citep{lin_sciscinet_2023} has opened the possibility for larger-scale analyses of research output and outcome generated by NIH and NSF funding \citep{yang_unveiling_2024}. Yet, there are still two open questions: (i) limited attention has been paid to thoroughly study the ID of funded research output as the existing works mainly focused on funding proposals \citep{bromham_interdisciplinary_2016,nichols_topic_2014}, and (ii) studying a more complete set of funded research beyond individual funder level remains a gap \citep{yang_unveiling_2024}.

We address these gaps by identifying and comparing all funded and unfunded publications in the field of infectious disease research in the WOS. By regarding funded and unfunded research as two dynamic processes and portraying the development of their ID from a temporal network perspective, our work also addresses the frequently stressed limitation that existing analyses of IDR are often static \citep{rafols_diversity_2010,wagner_approaches_2011,wang_consistency_2020,Rousseau_knowledge_2019}. Lastly, categorising papers based on journal classification systems causes inaccuracies \citep{wagner_approaches_2011,rafols_knowledge_2014,Rousseau_knowledge_2019} and such systems are often too coarse to capture knowledge integration at finer levels \citep{rafols_how_2012}. Therefore, instead of using journal classification systems like the WOSSC, we conduct topic-level analysis based on a selected set of major infectious diseases. 

\section{Data and Methods}
\label{Data and Methods}
\subsection{Data source and extraction procedure}
\label{Disease Selection and Data Source}
Infectious diseases continue to be a major global health issue with significant impact on human society \citep{baker_infectious_2022}, for example the COVID-19 pandemic, and the threat of climate change upon human health and infections \citep{mora2022over}. IDR is required to address this multifaceted challenge \citep{wilcox2005emerging}. Knowledge integration across infectious diseases has already contributed greatly to clinical and research knowledge \citep{schwetz_extended_2019}. Following one of our co-author's work  \cite{head_allocation_2020}, we based our study on research into 34 major infectious diseases.\footnote{The authors of \cite{head_allocation_2020} selected a group of 37 prominent infectious diseases for their funding versus disease burden analysis. Our selection excluded sexually transmitted infections and enteric infections because they are a higher level concept than others. We also added consideration of causal relations: trachoma is caused by chlamydia trachomatis but chlamydia also happens to be in the set, so we excluded trachoma.} In addition here, we included coronavirus and diphtheria. The inclusion of coronavirus is due to its recent prominent role in the field of infectious disease research. The inclusion of diphtheria is because two other highly related terms, pertussis and tetanus, have already been included in the group (the DTP combination vaccine prevents against all three diseases and they are thus commonly considered together; the three diseases will be addressed by DTP onwards). This gives us 36 selected infectious diseases which were considered as the units of analysis in this paper. They account for \$70 billion infectious disease research funding (65\% of overall total) from G20 countries between 2000 and 2017 \citep{head_allocation_2020}, and 94\% of the total infectious disease burden\footnote{Measured in Disability-Adjusted Life Years (DALYs). One DALY represents the loss of the equivalent of one year of full health.} in the 2021 Global Burden of Disease data\footnote{Available from \url{https://vizhub.healthdata.org/gbd-results/}. (Date last accessed: 2024.9.4)}. 

We selected the Web of Science (WoS) Core Collection database as our data source due to its extensive usage in the scientometrics studies \citep{birkle_web_2020}. 
The collection of publications on each disease can be seen as a body of knowledge \citep{hamburg2008considerations}, and thus the co-occurrence of diseases in the title, abstract or author keywords represents the practice of knowledge integration \citep{NAP11153}. We extracted, for each of the 36 diseases, both \textit{funded} research and \textit{all} research, the number of records containing the disease in their title, abstract, or author keywords, i.e., the number of occurrences. We also extracted, for both \textit{funded} and \textit{all} research, the number of records containing each possible pair of diseases in their title, abstract, or author keywords, i.e., the number of co-occurrences. 

To retrieve maximum funded records efficiently, we adopt the method of the right-hand truncation search strategy in both the funding agency (FO) and funding grants (FG) field suggested in \cite{liu_funding_2020}, as detailed in Search Query 1. We refer to papers that have FO or FG information in the WoS as funded research (labelled ``F")\footnote{\textit{All} research will be labelled ``A" hereafter.} and those that don't as unfunded research (labelled ``U"). We searched only the document types of original research article, data paper, and proceeding paper, and we specifically excluded any withdrawn or retracted publications, as detailed in Search Query 2. The timespan considered in this paper is from 1995 to 2022 inclusive\footnote{The WoS did not include the author keywords for articles until 1991. The years 1991-1994 were excluded from our analysis as the link profiles of these years were very volatile and did not seem to form coherent clusters with any of the rest of the years like the ones in \fref{Fig1}.}. See \tref{D36} for the search strategy for each disease\footnote{Search terms for each disease are firstly picked based on experts' advice and the entry terms (a list of synonyms) in the Medical Subject Heading (MeSH) database. A further selection of search terms was done after testing the added benefit of the search terms, i.e., search terms with low or no additionally identified papers were dropped.}\footnote{The search term of herpes in our study contains infections related to Herpes Simplex Viruses (HSV) and Shingles.} as well as the total and funded number of records returned.

\textbf{Search Query 1:} FO = (A* OR B* OR C* OR D* OR E* OR F* OR G* OR H* OR I* OR J* OR K* OR L* OR M* OR N* OR O* OR P* OR Q* OR R* OR S* OR T* OR U* OR V* OR W* OR X* OR Y* OR Z* OR 0* OR 1* OR 2* OR 3* OR 4* OR 5* OR 6* OR 7* OR 8* OR 9*) OR FG = (A* OR B* OR C* OR D* OR E* OR F* OR G* OR H* OR I* OR J* OR K* OR L* OR M* OR N* OR O* OR P* OR Q* OR R* OR S* OR T* OR U* OR V* OR W* OR X* OR Y* OR Z* OR 0* OR 1* OR 2* OR 3* OR 4* OR 5* OR 6* OR 7* OR 8* OR 9*)

\textbf{Search Query 2:} DT = ((Article OR Data Paper OR Proceedings Paper) NOT (Retracted Publication OR Withdrawn Publication OR Retraction))

To explore the possibility that an instance of co-occurrence of diseases in the abstract of a paper is not because they are the primary research objective of the paper but merely a mention as an aside we carried out some manual checks on a sample of papers. For this purpose, we selected five pairs of representative infectious diseases (HIV-TB, Dengue-Zika, Tetanus-Diphtheria, HCV-HBV, Clamydia-Gonorreahea) which have important contributions to the system dynamics and performed a sample test of size 100 on each pair to examine the proportion of wrong identifications. As shown in \tref{false positive}, we found that the false positive rates are consistently less than 3\% and we found no bias towards particular disease pairs.

The extracted occurrences and cooccurrences of infectious diseases were recorded in tensors $(C_F)_{36\times36\times28}$ for F and $(C_A)_{36\times36\times28}$ for A. An entry $c_{ij}^t$ of $C_F$ (or $C_A$) gives the cooccurrence count between disease $i$ and $j$ in F (or A) in time slice $t$\footnote{The 28 time slices correspond to years from 1995 to 2022.} if $i\neq j$, and the occurrence of $i$ otherwise. The tensor of U $(C_U)_{36\times36\times28}$ is a direct result of element-wise subtraction between $C_A$ and $C_F$. The data extraction was performed using the Web of Science API Lite\footnote{The Web of Science API Lite: support search and data integration using Web of Science data returned as JSON or XML. \url{https://developer.clarivate.com/apis/woslite} (Date last accessed: 2024.9.4)} in Python, and the network analysis was performed using R. For abbreviations of diseases used in this study, please refer to \tref{abbr}.

We used one of the most widely-used normalisation measures in scientometrics to normalise the cooccurrence matrix, namely the cosine \citep{eckHowNormalizeCooccurrence2009} (also known as the Ochiai index \citep{zhou_normalization_2016} or Salton's index \citep{adnani_similarity_2020}), i.e., $w_{ij}^{t} = \frac{c_{ij}^{t}}{\sqrt{c_{ii}^{t}c_{jj}^{t}}}$. In our setting, the cosine normalises the cooccurrences between $i$ and $j$ at time slice $t$ by the geometric mean of the occurrences of $i$ and $j$ at $t$ when $i\neq j$, and thus measures the correlation between numbers of publications about diseases. When $i=j$, $w_{ij}^{t}=0$. By normalising $C_A$, $C_F$ and $C_U$, we ended up with $W_A$, $W_F$, $W_U$, based on which we then performed the following analysis. Note that we regarded the collection of all WoS publications related to infectious diseases as the system of infectious disease knowledge, where $w_{ij}^{t}$ is the association, or the cognitive proximity, or the current extent of knowledge integration between disease $i$ and $j$ at time $t$. A high value of $w_{ij}^{t}$ implies an established knowledge link between the pair while a low value suggests a non-established state.


We also note an important limitation of using the WoS for funding analysis: funded research is under-represented in the database. The WoS has begun to collect funding information since Aug 2008, and the assignment of funding information of publications before this time has been done retrospectively \citep{liu_funding_2020}, so likely a lesser proportion of funded papers have been identified pre-2008 compared to post-2008 as can be seen from \tref{Fperc}. However, we can see research on infectious diseases has much better coverage of funding information than the average level pre-2008 publications (around three times more), which enabled us to perform analysis on pre-2008 years. We acknowledge that the error rate of pre-2008 analysis would tend to be relatively larger, but we argue that the effect of this should not be substantial: we focus on correlations and not absolute numbers – unless there is a bias in the un-identified funded research pre-2008, this should only cause noise and not systematically affect our results. In addition, the research practices of scientists could also be an underlying factor that has made funded research less identifiable \citep{alvarez-bornstein_funding_2017}, which highlights the need to better regulate research and funding practices. 

\subsection{Network measures and analysis}
\label{Network Measures for ID}
We defined a temporal network to model the evolution of the ID of infectious disease research, with the nodes being the bodies of infectious disease knowledge and links being their connection strength quantified by the cosine. To quantify ID, we studied the coherence and intermediation measures of the constructed networks.\footnote{Note that every measure discussed in this section is computed given a time $t$, so we omitted index $t$ for simplicity.} 

We adopt a similar measure of coherence with \cite{rafols_diversity_2010}, i.e., the mean linkage strength, due to its simplicity, although at a different level of aggregation. We used the mean node strength to represent the overall compactness of the knowledge system, i.e.,
\begin{equation}
    \overline{s} = \frac{1}{N}\sum_is_i = \frac{1}{N}\sum_i\sum_{j\neq i}w_{ij}
\end{equation}
where $N$ represents the number of nodes in the network. Note that the node strength $s_i$ is a measure of how well an infectious disease integrates knowledge locally. 

Following \cite{leydesdorff_betweenness_2007}, we adopted betweenness centrality (BC) \citep{brandes_faster_2001} to track node-level ID from the intermediation perspective, i.e., one disease's ability to bridge otherwise disjoint disease knowledge. BC of a node $i$ is computed as 
\begin{equation}
    b_i = \sum_{j\neq k}\frac{g_{jk}(i)}{g_{jk}}
\end{equation}
where $g_{jk}$ is the number of shortest path between any two nodes and $g_{jk}(i)$ is the number of shortest path between two nodes going through node $i$. 

\cite{soos_beyond_2012} proposed network diameter (or the maximal shortest path length) as a measure of both coherence and intermediation. This measure provides an upper bound of how quickly information flows from one place to another. We proposed a similar measure, the average shortest path length (ASPL) \citep{jahanshadGeneticsPathLengths2012}, to capture the average shortest distance between infectious diseases. A low ASPL indicates that the disease network is compact. The shortest path length \citep{brandes_faster_2001} between node $i$ and $j$ is 
\begin{equation}
    d(i,j) = \min\left(\frac{1}{w_{ih}}+\cdots+\frac{1}{w_{hj}}\right)
\end{equation}
with $h$ being intermediary nodes between node $i$ and $j$. ASPL is the average of $d(i,j)$ over all possible pairs of diseases $(i,j)$.

Closeness centrality \citep{freeman2002centrality} offers a comparative perspective to betweenness centrality. As a node-level measure, it quantifies how close a disease is to all other diseases in the network. It is computed as
\begin{equation}
    c_i = \frac{1}{\frac{1}{n}\sum_{j=1}^{n}d(i,j)}
\end{equation}
where $n$ is the number of all reachable node from $i$.

Beyond coherence and intermediation measures proposed in earlier literature, we were also interested in clustering patterns of the infectious disease knowledge network. For this purpose, following \cite{blondel_fast_2008}, we evaluated the networks modularity and used the Louvain method \citep{blondel_fast_2008} to determine modules of interconnected diseases. Modularity is computed as 
\begin{equation}
    Q = \frac{1}{2m}\sum_{i,j}\left(w_{ij}-\frac{s_is_j}{2m}\right)\delta(c_i,c_j)
\end{equation}
where $m$ is the sum of link weights, $s_i$ refers to the strength of node $i$, and $\frac{s_is_j}{2m}$ is the expected link strength between $i$ and $j$ assuming a random distribution of connections which preserves the strength distribution across nodes. A high modularity $Q$ represents a well-defined community structure with many intra-community links and few links connecting separate communities, i.e., compartmentalisation \citep{cohen2016segregation}, while a low $Q$ indicates a weak community structure.

\section{Results}
\label{Results}
\subsection{RQ1: Identification of temporal regimes in interdisciplinary research}
\label{Formation of infectious disease regimes}
\begin{figure}[htp]
    \centering    
    \includegraphics[width=\textwidth]{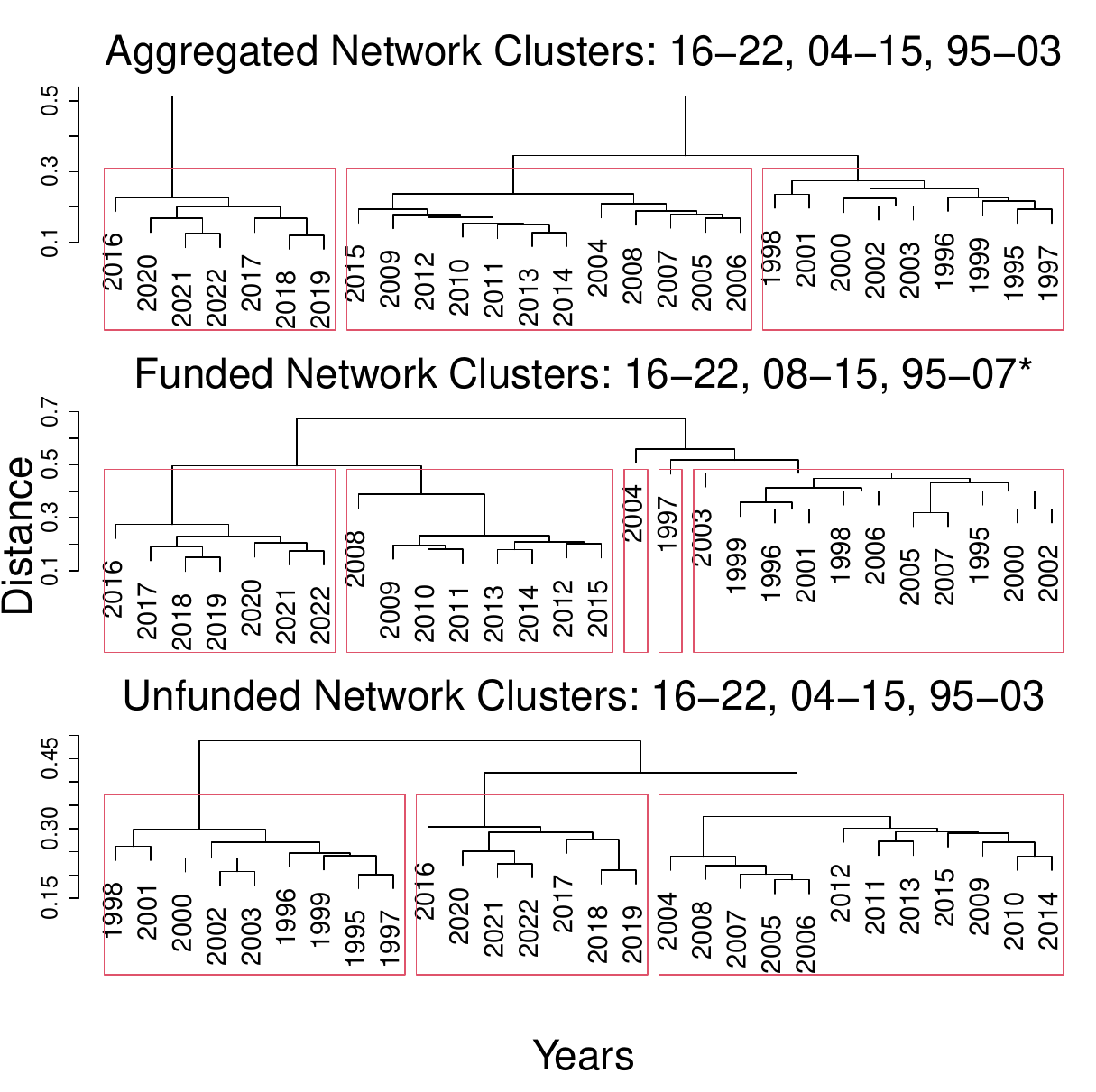}
    \caption{Bottom-up hierarchical clustering of yearly time slices for the funded, unfunded and aggregated networks. Hierarchical clustering using UPGMA was performed for all (A), funded (F), and unfunded (U) networks respectively. The highlighted identified significant clusters were based on the elbow method as in \fref{Fig8} in the appendix.}
    \label{Fig1}
\end{figure}

Given two yearly slices of the correlation network at times $t_1$ and $t_2$, $w^{t_1}$ and $w^{t_2}$, we can quantify the distance between them by the Euclidean distance, i.e., $\sqrt{\sum_{i\neq j} (w_{ij}^{t_2}-w_{ij}^{t_1})^2}$. We then used bottom-up hierarchical clustering based on UPGMA (Unweighted Pair Group Method with Arithmetic mean) \citep{Sokal1958ASM} to partition the temporal evolution of publication patterns into distinct regimes. Further analysis of the obtained partitioning using the elbow method reveals three regimes for funded, unfunded and all research respectively (see \fref{Fig1}) which we identify with coherent periods of time, i.e. 1995-2007 (F1)\footnote{We combined the single element clusters 1997 and 2004 with the bigger cluster to make F1 continuous as all years within F1 are already quite distant compared with that of F2 and F3. Years in F1 being more distant might be attributed to the larger noise in funded research as it is under-represented pre-2008 in the WoS, as discussed in \sref{Disease Selection and Data Source}. Visualisations with and without 1997 and 2004 were compared in \fref{Fig7}.}, 2008-2015 (F2), and 2016-2022 (F3) as regimes for funded research; 1995-2003 (U1), 2004-2015 (U2) and 2016-2022 (U3) for unfunded research; 1995-2003 (A1), 2004-2015 (A2), and 2016-2022 (A3) for all research\footnote{Note that unfunded research has the same regimes as all research, so the following analysis will only be focusing on funded and unfunded regimes.}.  

\begin{figure}[htp]
    \centering
    \includegraphics[width=\textwidth]{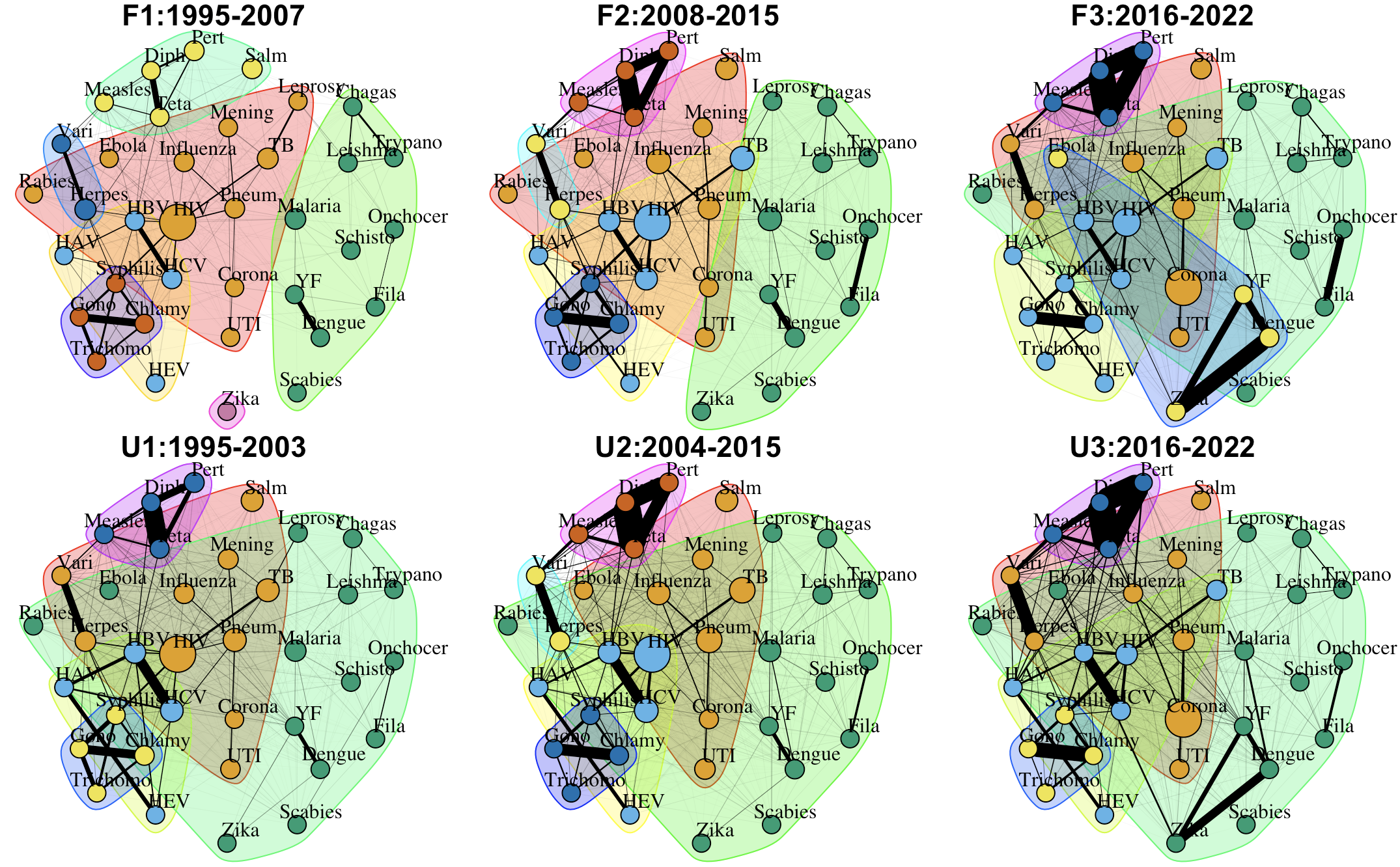}
    \caption{Network visualisation of funded and unfunded temporal regimes. The link strength $w_{ij}$ shown in the visualisation of the regime is computed by the average of the link strength across the regime. Node size of disease $i$ represents the proportion of publications on $i$ during the regime. The network layout was produced using R based on the Fruchterman-Reingold method \citep{frlayout} applied on F1. The community structure was detected based on the Louvain method \citep{blondel_fast_2008}.}
    \label{Fig2}
\end{figure}

Regimes for F and U were visualised in \fref{Fig2}. Each visualisation represents the average network in the corresponding regime. We observe regimes F1 and U1 tend to be relatively weakly connected compared with later regimes, indicating a weaker level of knowledge integration. This can also be seen from \tref{global_measure} where the average node strength $\overline{s}_F$ and $\overline{s}_U$ in F1 and U1 are 0.24 and 0.35, the lowest across all regimes. We note the existence of a well-integrated community of four curable sexually transmitted infections (STIs), i.e., chlamydia, gonorrhoea, syphilis and trichomoniasis in F1 and U1. The following links in F1 and U1 also demonstrate strong extents of integration: Hepatitis B (HBV) and Hepatitis C (HCV), dengue and yellow fever, varicella and herpes\footnote{Varicella is linked with shingles as they are both caused by Varicella-Zoster Virus (VZV).}, and diphtheria and tetanus (especially in U1). 

Investigating the 2nd temporal regime, we observe that both F2 and U2 are more densely connected compared with the previous regimes, as can also be seen from $\overline{s}_F$ and $\overline{s}_U$ rising to 0.35 and 0.43, respectively, see \tref{global_measure}. What stands out is that the knowledge integration among the three diseases of DTP is strongly reinforced. 

Moving to F3 and U3, we observe that DTP gets integrated further, and the group of vector-borne (Aedes mosquitoes) diseases, i.e., yellow fever, zika and dengue, becomes relatively well integrated. Throughout the entire time period, the continuous integration of DTP knowledge is very likely caused by an increase in vaccination-related research. The sudden increase in the integration between zika, yellow fever, and dengue in the last regime is highly likely due to the increased attention on zika and its closely related neighbours on the knowledge network, as a reaction towards the emergence of zika as a public health emergency in 2016\footnote{On 01 Feb 2016, WHO declared Zika and its complications constitutes a Public Health Emergency of International Concern. See the discussion for further details.}. The integration of the four curable STIs (chlamydia, gonorrhoea, syphilis and trichomoniasis) and the group of hepatitis infections is found to be relatively stable, which might indicate an already well-established knowledge structure in those sub-fields.

We measure individual disease contributions to system change by characterising each disease's relative impact in terms of relative link strength change $r_{s,within}$ versus the volatility $\sigma(diff[s_i(t)])$ in its time evolution of link strengths. More precisely, we defined $\sigma(diff[s_i(t)])$ to be the standard deviation of the change in node strength $s_i$ between consecutive years in the regime ($diff$ represents differencing of the time series), and $r_{s,within}$ to be the change of node strength from the beginning to the end of a regime normalised to the average level of change of all nodes, i.e., if a regime starts from year $t_1$ and ends in year $t_2$, $r_{s,within}=\frac{\overline{s_{i,t_2}}-\overline{s_{i,t_1}}}{\left|s_{t_2}-s_{t_1}\right|}$, where $\overline{s_{i,t_2}}$ represents the mean of $s_i$ across $t_2$, $t_2-1$, $t_2-2$, $\overline{s_{i,t_1}}$ represents the mean of $s_i$ across $t_1$, $t_1+1$, $t_1+2$, and $s_{t_2}$ and $s_{t_1}$ represents the average of $\overline{s_{i,t_2}}$ and $\overline{s_{i,t_1}}$ over all $i$ respectively. An increase in $r_{s,within}$ indicates the presence of relatively stronger local connections, thus being more integrated locally during the regime. In \fref{Fig3}, we plotted the relative link strength change $r_{s,within}$ on the y-axis versus the volatility $\sigma(diff[s_i(t)])$ on the x-axis. We observe that the changes in the different regimes seem to be driven by different diseases (\fref{Fig3}). Hepatitis A (HAV), chlamydia and dengue gain relatively more strength in F1 while yellow fever and three curable STIs (gonorrhoea, syphilis, and trichomoniasis) stand out in U1 (\fref{Fig3}). DTP emerges during F2 while varicella and pertussis stand out during U2. Throughout F3, there is a significant gain in the strength of coronavirus and zika, while for U3, coronavirus, tetanus and influenza stand out. All of the highlighted infectious diseases emerge with rather high volatility. The strong emergence of coronavirus-related bodies of knowledge in F3 and U3 is likely due to the huge attention paid to the COVID-19 global pandemic since 2019.

\subsection{RQ2: Compartmentalisation and integration of funded and unfunded research}
\label{Trends in compartmentalisation and integration of funded and unfunded research}
In \tref{global_measure} we reported the average node strength $\overline{s}$, the modularity $Q$ and the average shortest path length $ASPL$ of the averaged networks for the regimes identified in \sref{Formation of infectious disease regimes}. Comparing trends in $\overline{s}$, $Q$ and $ASPL$ for the different time regimes we make the following observations. First, regarding the average node strength $\overline{s}$, we find $\overline{s}_F$ increases from 0.24 to 0.35 to 0.44 and $\overline{s}_U$ from 0.35 to 0.43 to 0.54. This increase in the intensity of connection indicates that funded and unfunded research has become more interdisciplinary. Second, regarding the average shortest path length $ASPL$, we note a substantial decrease over time. The observed trend indicates a tendency towards higher interdisciplinarity in funded and unfunded research through stronger intermediation. Third, regarding the modularity $Q$, we find an opposite trend with $Q_F$ increasing and $Q_U$ decreasing, meaning that funded research becomes more compartmentalised while unfunded research becomes more globally integrated. 

From the above, we infer that knowledge integration in funded research tends to reinforce community structure. This indicates that funded research tends to be more specialised and stays on the conservative side, i.e., it focuses on deepening already established relationships of diseases \citep{rzhetsky2015choosing,foster2015tradition}. Knowledge integration in unfunded research, on the other hand, tends to take place through weakening community structures, which indicates that unfunded research is relatively less conservative and focuses more on bridging distant diseases\footnote{The observation that unfunded research is more widely exploratory in terms of ID is consistent with the fact that unfunded researchers have a higher degree of freedom in setting their own goals \citep{edwards_why_2022}.}. These interpretations are further supported by results shown in \fref{Fig9} and \fref{Fig10}, where we have compared relative increments in link strengths between time periods for the strongest (top 5\%) and weaker (bottom 50\%) links in the evolution of IDR in funded and unfunded research. We note, that the strongest links in funded research have consistently gained more in link strength compared to unfunded research (relative gains $4.96>3.74$ from 1995-2008 to 2009-2015 and even more so $4.13>2.67$ from 2009-2015 to 2016-2022). In contrast, for the majority of weaker links, links have gained more in strength for unfunded research than for funded research (relative gains $0.26>0.16$ from 1995-2008 to 2009-2015 and $0.22>0.15$ from 2009-2015 to 2016-2022). Both observations make it very clear that strength gains in funded research tend to be more aligned with already strong connections, whereas strength gains in unfunded research tend to be more exploratory, reinforcing weak connections.

\begin{table}[htp]
    \centering
    \begin{tabular}{||c||c|c|c||}
    \hline\hline
    Periods  & F1:1995-2007 & F2:2008-2015 & F3:2016-2022 \\
    \hline
    $\overline{s}_F$& 0.24 & 0.35 & 0.44 \\
    \hline
    $Q_F$ & 0.46 & 0.47 & 0.49\\
    \hline
    $ASPL_F$ & 127 & 98 & 89\\
    \hline\hline
    Periods  & U1:1995-2003 & U2:2004-2015 & U3:2016-2022 \\
    \hline
    $\overline{s}_U$& 0.35 & 0.43 & 0.54 \\
    \hline
    $Q_U$& 0.45 & 0.43 & 0.40\\
    \hline
    $ASPL_U$& 92 & 77 & 64\\
    \hline\hline
    \end{tabular}
    \caption{Average node strength $\overline{s}$, modularity $Q$ and average shortest path length $ASPL$ for funded (F) and unfunded (U) regimes.}
    \label{global_measure}
\end{table}

\subsection{RQ3: Analysing research funding into Interdisciplinary Research}
\label{Analysing research funding into IDR}
In \sref{Trends in compartmentalisation and integration of funded and unfunded research}, observing that in contrast to unfunded research funded research tended to become more compartmentalised, we noted different trends in the organisation of funded and unfunded research over time. Here, we are interested in a more complete understanding of how research investment has driven the evolution of interdisciplinarity. For this purpose, one could see unfunded research as representing general scientific interest and compare this to research driven by funding allocation. 

\begin{figure}[htp]
    \centering
    \includegraphics[width=\textwidth]{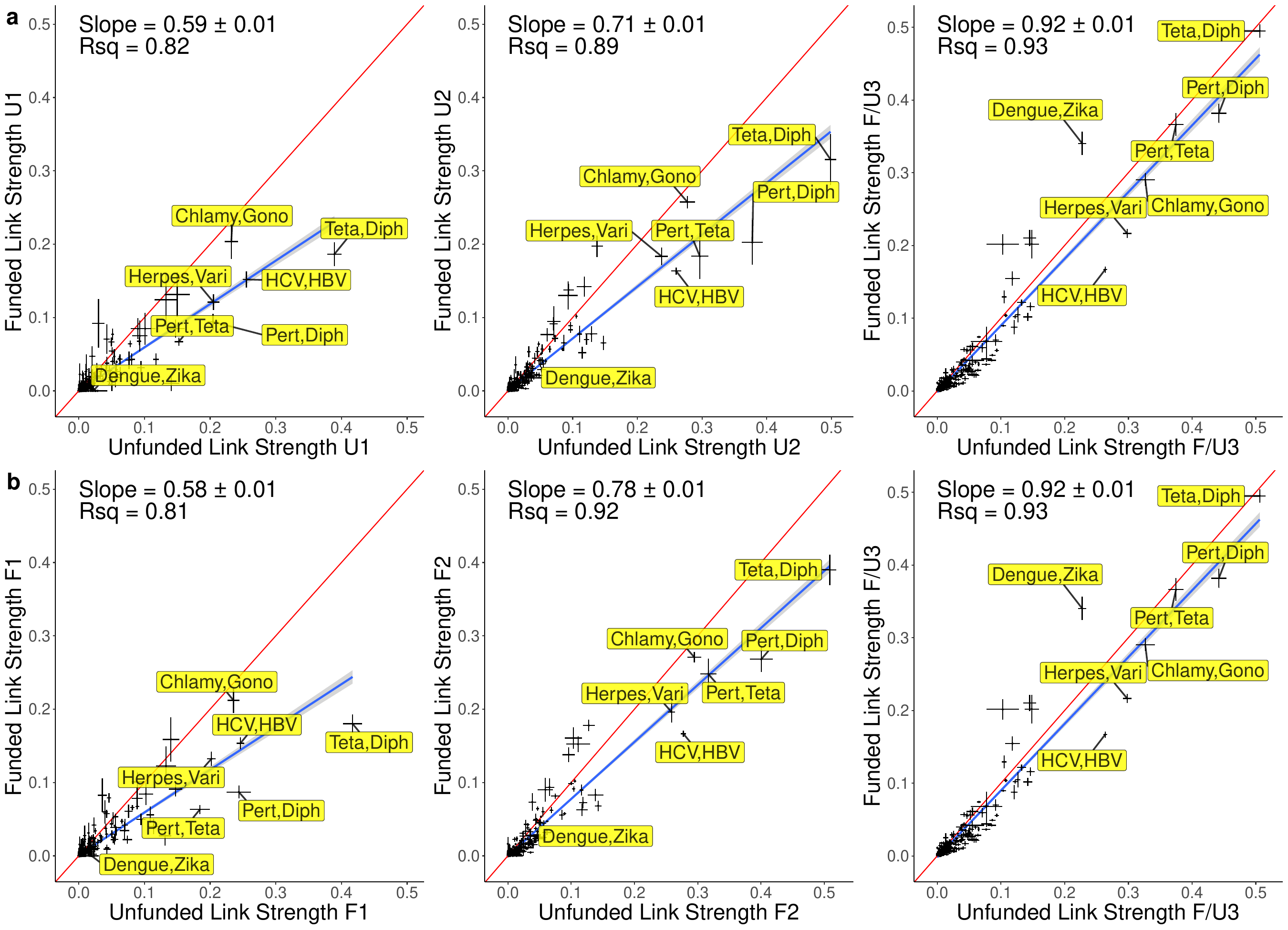}
    \caption{The evolution of the level of knowledge integration in funded research compared with unfunded research for (A) U1:1995-2003, U2:2004-2015 and U3:2016-2022; and (B) F1:1995-2007, F2:2008-2015 and F3:2016-2022. For each plot, the y-axis represents the average link strength of a pair of infectious diseases in funded research and the x-axis in unfunded research. Each error bar represents the standard error of the link strength of a pair within the regime. The red line represents the 45-degree line: any link lying on the line represents the same level of ID in funded and unfunded research, above (below) indicates more (less) funding is allocated to the pair than should be. A regression line is fitted to the points and the slope of the fitted regression line, the standard error of the slope, and the R-squared value of the fitted regression line were reported in the top left corner. The shaded area around the fitted line represents one standard error of the slope. Yellow labels represent the top seven influential links that appeared in F3 or U3, and their corresponding positions were shown in F2, F1, U2, and U1 plots.}
    \label{Fig4}
\end{figure}

To operationalise this comparison, for each temporal regime, we measured average correlations between pairs of diseases in funded and unfunded research and plotted them against each other in \fref{Fig4}. Note that in \fref{Fig4} we show results based on partitions for the temporal regimes in both U and F (top and bottom rows), which show essentially the same trends. The 45-degree line in \fref{Fig4} represents a situation where research corresponding to a pair of diseases is as well-funded as represented in general scientific interest, i.e. in unfunded research. Inspecting the figure, we first observe that most of the pairs stay below the 45-degree line, indicating that the level of interdisciplinarity in funded research is generally lower than in unfunded research. In other words, IDR in infectious disease research tends to be underfunded. However, this changes over time as the slope of the fitted regression line keeps increasing and approaches one, getting closer to where funding allocation matches overall scientific attention. Simultaneously, also the R-squared value increases over time, indicating an increasingly closer alignment between research investment and scientific interest regarding the importance of interdisciplinary infectious disease areas.

Further to the above, \fref{Fig4} also allows to analyse how well a pair of diseases is funded relative to all other pairs. Visually, such relative over- or underfunding is indicated by whether the corresponding datapoint is above or below the fitted regression line. For instance, in \fref{Fig4}, we highlight seven pairs of diseases that received the most scientific attention in the last temporal regime. Looking at trends over time, we note that in F1 and U1, the research area chlamydia-gonorrhoea stays above the fitted regression line indicating a relatively well-funded status, whereas the DTP-related areas stay slightly below the line indicating they are relatively underfunded. These observations remain true for F2 and U2, except that HCV-HBV becomes slightly underfunded and DTP-related areas move slightly closer to the fitted line. In F3 or U3, dengue-zika becomes very well-funded relative to other areas whereas HCV-HBV becomes relatively underfunded. 



\subsection{RQ4: Comparing the roles of diseases in knowledge integration} 
\label{Identifying leaders and drivers of knowledge integration}
To identify the diseases with the strongest global and local impact, we ranked the diseases by strength, betweenness and closeness. \tref{Node_measure} reports the top three ranked strengths $s_i$, betweenness centrality $b_i$ and closeness centrality $c_i$ of the infectious diseases for F and U regimes. $s_i$ represents the knowledge integration at a local level and $b_i$ and $c_i$ at a global level, thus comparing these three allows us to compare the local and global impact of the top-ranked diseases. Comparing the different time regimes we make the following observations. Regarding the betweennees centrality $b_i$ and closeness centrality $c_i$, we find HIV has the highest $b_i$ and $c_i$ in all regimes for both F and U, followed by other prominent diseases like tuberculosis, HBV, or malaria. This indicates that prominent infectious diseases like HIV stay in the centre of the network and bridge distant bodies of knowledge. Regarding $s_i$, we find DTP are the top three in $s_i$ in F2, F3, U2 and U3 with a remarkable increase in magnitude through time, but none of DTP appears in the top three $b_i$ of F or U regimes. This indicates that DTP stay on the periphery of the network and only enhance established connections between bodies of knowledge.

\begin{table}[htp]
    \centering
    \begin{tabular}{||c||c|c|c||}
    \hline\hline
    Regimes  & F1:1995-2007 & F2:2008-2015 & F3:2016-2022 \\
    \hline\hline
    Top Three $s_i$   & \Centerstack{ HIV(0.59) \\Gono(0.51)\\Chlamy(0.51)} & \Centerstack{Teta(0.89)\\Diph(0.86)\\Pert(0.71)} & \Centerstack{Teta(1.18)\\ Diph(1.17)\\ Pert(1.03)} \\
    \hline
    Top Three $b_i$ & \Centerstack{ HIV(0.44)\\ TB(0.24) \\ Leprosy(0.20) }& \Centerstack{HIV(0.41) \\ Malaria (0.36)\\HBV(0.18) }& \Centerstack{HIV(0.32) \\Malaria(0.24)\\Pneum(0.15) }\\
    \hline

    Top Three $c_i$ & \Centerstack{ HIV(0.013)\\ TB(0.012) \\ HCV(0.012) }& \Centerstack{HIV(0.016) \\ HCV(0.015)\\ Syphilis(0.015) }& \Centerstack{HIV(0.016) \\HCV(0.015)\\TB(0.015) } \\

    \hline\hline
    Regimes  & U1:1995-2003 & U2:2004-2015 & U3:2016-2022 \\
    \hline\hline
    Top Three $s_i$ &\Centerstack{Diph(0.86)\\ Teta(0.83)\\ HBV(0.82)} & \Centerstack{Diph(1.26)\\ Teta(1.19)\\ Pert(0.99)} & \Centerstack{Diph(1.46)\\ Teta(1.41)\\ Pert(1.32)} \\
    \hline
    Top Three $b_i$ & \Centerstack{HIV(0.30) \\ HBV(0.29)\\Syphilis(0.20) }& \Centerstack{HIV(0.27) \\Malaria(0.22) \\ TB(0.16)}& \Centerstack{HIV(0.25)\\ Malaria(0.17)\\ Dengue(0.16)}\\
    \hline

    Top Three $c_i$ & \Centerstack{ HIV(0.017)\\ HBV(0.016) \\ HCV(0.016) }& \Centerstack{HIV(0.020) \\ TB (0.019)\\HBV(0.018) }& \Centerstack{HIV(0.023) \\Measles(0.022)\\Diph(0.022) } \\

    \hline\hline
    \end{tabular}
    \caption{Top three ranked diseases, ranked by node strength $s_i$, betweenness centrality $b_i$, and closeness centrality $c_i$ for the funded (F) and unfunded (U) regimes.}
    \label{Node_measure}
\end{table}

We investigated the relative change in the diseases' betweenness $r_{b, within}$ with respect to their volatility $\sigma(diff[b_i(t)])$ in \fref{Fig5}. 
$r_{b, within}$ and $\sigma(diff[b_i(t)])$ were calculated in analogy to $r_{s, within}$ and $\sigma(diff[s_i(t)])$ in \sref{Formation of infectious disease regimes}. Note that $r_{b, within}$ for each infectious indicates a change in its ability to act as a bridge between other diseases relative to the average change in betweenness of all infectious diseases within a regime. From \fref{Fig5}, we note that HCV, HAV and HIV have gained notable betweenness during F1, while dengue and HAV have gained the most betweenness during U1. Malaria has gained strongly in betweenness in F2 with high volatility, whereas during U2 pertussis and varicella have gained betweenness with moderate volatility. Throughout F3, coronavirus has gained significant betweenness with moderate volatility, while during U3 it is ebola, pertussis and measles that have gained notable betweenness. The significant gain of coronavirus is likely due to the COVID-19 pandemic, but interestingly this pattern is not found in unfunded research. 

\subsection{RQ5: Quantifying the Role of Coronavirus Research in IDR}
\label{Quantifying the Role of Coronavirus Research in IDR}
\begin{figure}[htp]
    \centering
    \includegraphics[width=\textwidth]{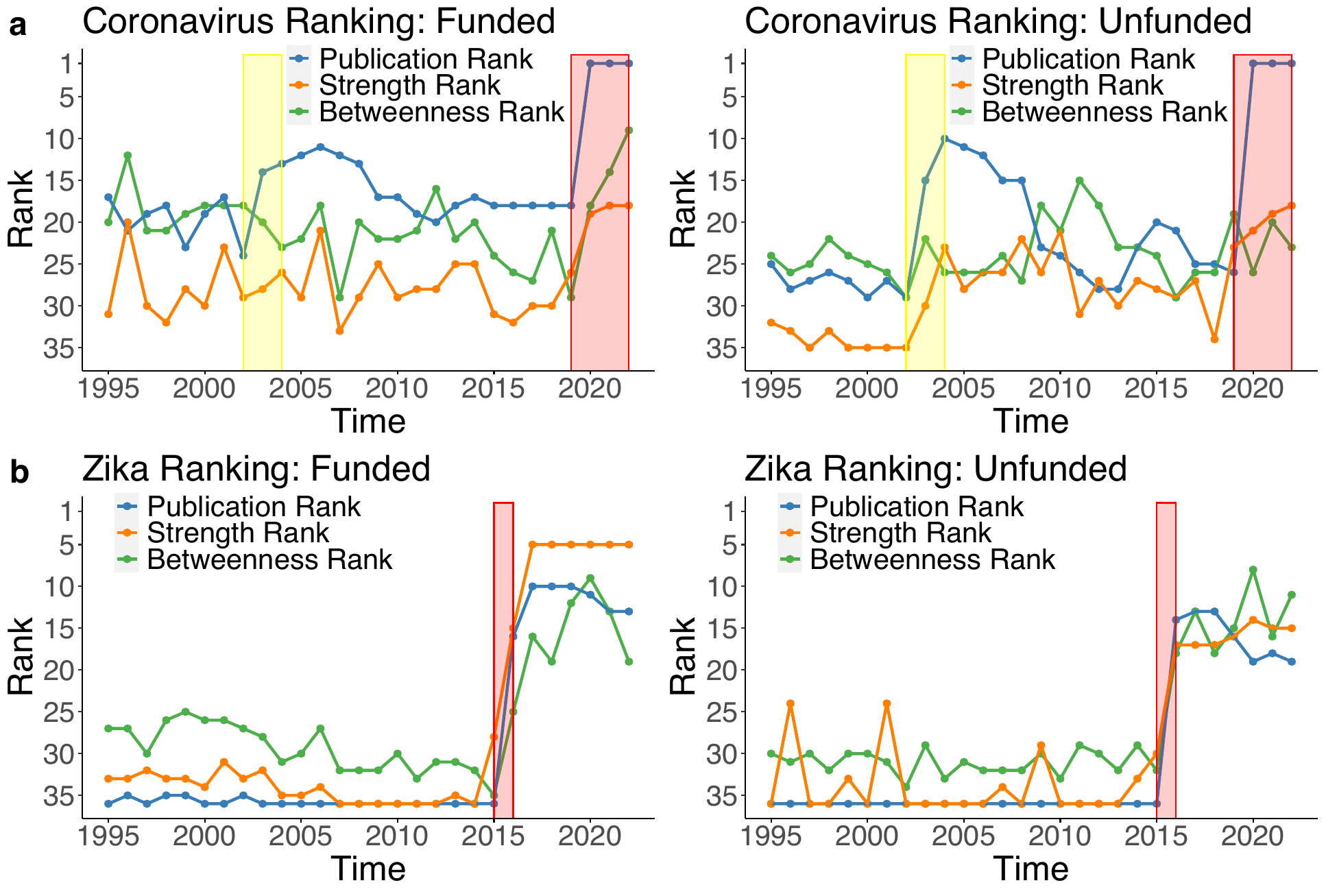}
    \caption{Temporal change in the ranking of the annual number of publications, strength and betweenness of (a) coronavirus-related publications (b) zika-related publications in the funded and unfunded disease network 1995-2022. The yellow area in (a) indicates the SARS outbreak 2002-2004, and the red area in (a) indicates the COVID-19 outbreak 2019-2022. The red area in (b) indicates the zika outbreak 2015-2016.}
    \label{Fig6}
\end{figure}

To further explore the role of the coronavirus pandemic in infectious disease research, in \fref{Fig6}(a) we reported the changing ranking of coronavirus research in terms of the number of publications, node strength and betweenness over time. We also marked two important coronavirus-related public health events in the figure, i.e., the 2002-2004 SARS outbreak (yellow regions) and the 2019-2022 COVID-19 pandemic (red regions). 

For both the SARS and COVID-19, we observe a significant rise in the publication ranking of both F and U. During SARS the publication ranking of coronavirus broke into the top 15 for F and top 10 for U, while during COVID-19 it attained the top rank for both F and U. Unsurprisingly, we see that both events led to an increase in scientific attention, with COVID-19 to the greatest extent. This is consistent with the result in \fref{Fig2} where the proportion of coronavirus-related research (represented by the node size) outweighs HIV and becomes the highest in F3 and U3. However, in terms of the ranking of node strength, despite experiencing moderate ranking gains during SARS and COVID-19, coronavirus stayed out of the top 15 for both F and U. This might indicate that the outbreaks caused some local knowledge integration around coronavirus but not at a significant extent at the system level. 

Moreover, in terms of the betweenness ranking, U exhibited only minor fluctuations around the 20th to 25th during both events, while F showed a minor drop around the 20th rank during SARS but a considerable jump from the 30th to a top 10 position during COVID-19. Such a jump might indicate coronavirus has become increasingly important by moving more into the centre of the infectious disease network and starting to bridge distant knowledge. However, even though coronavirus has had a very important role in terms of the number of publications, its systemic impact on the interdisciplinarity of infectious disease research has been relatively small to date. This might be due to the fact that coronavirus is conceptually not so strongly related to other disease areas. Another potential reason is there perhaps exists a temporal delay before systemic impact is observed. 

To explore the potential for delays in the systemic impact of COVID-19 on coronavirus research, we compared to another infectious disease with a sudden increase in prominence for which a longer timeframe of observations is available. This is provided by the zika virus which strongly gained in attention during the global outbreak between 2015 and 2016. For both F and U, strength seemed to immediately follow publication ranking during the outbreak 2015-2016 (\fref{Fig6}(b)), which is different to COVID-19. The betweenness ranking in F lagged a bit behind but then became quite volatile, while in U there seemed no delay. We conclude there is no sufficient evidence for a delay in the systemic impact of the zika outbreak. Funded research of zika has demonstrated extraordinary local integration (perhaps with other vector-borne diseases) with strength ranking being the best performer among all three rankings (peaked at the top 5 and sustained at that level ever since). This result on funded research on zika shows consistency with results in \sref{Formation of infectious disease regimes}, \sref{Trends in compartmentalisation and integration of funded and unfunded research}, and \sref{Analysing research funding into IDR}.

\section{Discussion}
\label{Discussion}

We summarise the results and discuss the contribution of this study in \sref{Summary of results}, then discuss the implications of the results, limitations and future work in \sref{Real-world factors, limitations, and future works}.

\subsection{Summary of results and contributions}
\label{Summary of results}
We have investigated the evolution of interdisciplinarity in funded and unfunded research on infectious diseases over the period 1995-2022. Constructing correlation networks of research output relating to pairs of diseases, we identified three regimes for funded and unfunded research respectively where each regime is a coherent period of time characterised by a particular knowledge structure. Based on the regimes, we found that both the funded and unfunded research had an increase in the extent of knowledge integration through time in terms of coherence and intermediation. However, while increases in interdisciplinarity in funded research took place through compartmentalisation, increases in unfunded research have typically occurred through global integration. Besides, we also found that IDR on infectious disease is underfunded in general but also note that this effect has decreased through time. Our analysis further allows to identify individual well-funded and underfunded interdisciplinary areas. 

Investigating the role of individual diseases in these trends, we found IDR on prominent diseases like HIV, malaria and tuberculosis has strong bridging effects, while IDR on diphtheria, tetanus, and pertussis has strong local enhancement. Lastly, we found that coronavirus has attracted the most publications in infectious disease research since the emergence of COVID-19. In spite of this, however, the systemic impact of coronavirus research to date on infectious disease knowledge integration has been relatively small.

The results of this study contribute to the understanding of the roles played by the research on individual infectious disease in interdisciplinary knowledge generation, and the relationship between global public health emergencies and interdisciplinary research efforts. These could provide valuable insights for future health priority setting, for example with horizon scanning for new and emerging threats to health, such as pandemic planning.

Overall, this research presents a generalisable framework to examine the impact of funding in interdisciplinary knowledge creation and provides important insights to science policy.



\subsection{Implications, limitations and future work}
\label{Real-world factors, limitations, and future works}

Conservatism in scientific knowledge \citep{foster2015tradition,rzhetsky2015choosing} manifests in infectious disease research through a compartmentalized structure shaped by funding patterns. This trend aligns with broader observations by \cite{park_papers_2023} that scientific disruptiveness and innovation are declining due to an over-reliance on a narrower scope of existing knowledge—a "conservative trap" \citep{Fortunato_science_2018}. The conservatism in idea selection in science comes down to scientists' inclination to prefer productivity over riskier innovation (IDR for instance), where the rewards of additional impact do not compensate for the risk of publishing nothing \citep{Fortunato_science_2018}. Research institutions and funding agencies should encourage intelligent risk-taking through establishing mechanisms that diversify risks across a portfolio of scientific projects during evaluation \citep{rzhetsky2015choosing}, perhaps adopting models like the group-based evaluations at Bell Labs or the people-centred approach of the Howard Hughes Medical Institute.

Such conservatism could also be attributed to and reinforced by rigid disciplinary-based research evaluations \citep{Rousseau_knowledge_2019,woelert_paradox_2013,rylance2015global,fontana2022interdisciplinarity} where evaluators that are subject to bounded rationality and hold disciplinary-based standards tend to penalise novelty \citep{boudreau_looking_2016,packalen_nih_2020,woelert_paradox_2013}. As a consequence, despite being repeatedly advocated in science policy documents \citep{woelert_paradox_2013}, IDR becomes systematically disadvantaged \citep{bromham_interdisciplinary_2016}, forming the "paradox of interdisciplinarity" \citep{woelert_paradox_2013}. Addressing this might require a dual approach: relaxing stringent classification systems, despite the political and bureaucratic costs \citep{woelert_paradox_2013}, and directly channelling support to interdisciplinary initiatives \citep{lyall2013role} through schemes like the UKRI's cross research council responsive mode pilot scheme or the Gates Foundation’s Grand Challenges. Moreover, establishing interdisciplinary research centres, as the NSF has done, could further support this shift, although the impact of such centres should be continuously assessed to confirm their efficacy \citep{woelert_paradox_2013}.






The rising consistency between funded and unfunded research might suggest a better allocation of funding in the sense that interdisciplinary research areas with high scientific interest have been addressed by research funding. However, this would need further in-depth examination as we did not take into account the complex interplay between funded and unfunded research in our study. 

We note that the focus of our work is on topic-level ID instead of ID based on conventional classification systems. Looking at a specific field like infectious disease research and considering the sub-topics allows us to capture the more granular level dynamics of knowledge integration that has typically been overlooked by past research \citep{rafols_diversity_2010,rafols_knowledge_2014}. However, we note that the granularity of the choice of classifications is likely to have an impact on the resulting ID \citep{Rousseau_knowledge_2019} and we leave an in-depth examinations of the effects of granularity for future work. We also note that although this research focuses on the dynamics of infectious disease research, the proposed framework is generalisable to any other research field.


We distinguished (the research on) infectious diseases with two different patterns of knowledge integration: diseases that tend to integrate locally (e.g. DTP) and diseases that tend to integrate globally (e.g. HIV, tuberculosis, and malaria). There is no simple answer to what drives these integrations as the collection of knowledge on infectious diseases spans multiple domains including pathogenesis, diagnosis, cause, treatment, prognosis, spread, prevention, social impact and policy \citep{hamburg2008considerations}. However, there may be fundamentally two types of driving factors: intentional or unintentional \citep{wagner_approaches_2011,glanzel_various_2022}. Policymakers, funders and researchers have been targeting specific groups of diseases to meet their respective priorities and goals: these are intentional efforts, and they may tend to focus on more established groups of diseases that have been associated with great disease burdens at that time. For instance, the WHO has created global health programmes focusing on Hepatitis B and C\footnote{\url{https://www.who.int/teams/global-hiv-hepatitis-and-stis-programmes/hepatitis/overview} (Date last accessed: 2024.9.4)}, STIs\footnote{\url{https://www.who.int/teams/global-hiv-hepatitis-and-stis-programmes/stis/overview} (Date last accessed: 2024.9.4)}, and vector-borne infections\footnote{\url{https://www.who.int/publications/i/item/9789241512978} (Date last accessed: 2024.9.4)}. Therefore, it might be that emergence of the diseases with strong local integration is the result of such intentional effort. For example, DTP, the diseases found with the largest extent of local integration, have been a part of the WHO Expanded Programme on Immunization\footnote{\url{https://www.who.int/teams/immunization-vaccines-and-biologicals/essential-programme-on-immunization} (Date last accessed: 2024.9.4)}, 
and a focus of Gavi, the Vaccine Alliance\footnote{\url{https://www.gavi.org/news/media-room/gavi-helps-dtp3-coverage-rise-after-stagnation} (Date last accessed: 2024.9.4)}, and there has been ongoing development of a variety of DTP-related combination vaccines\footnote{\url{https://www.who.int/teams/health-product-policy-and-standards/standards-and-specifications/norms-and-standards/vaccine-standardization/dt-based-combined-vaccines} (Date last accessed: 2024.9.4)}.

Unintentional factors like discoveries on biological associations, patterns of comorbidity, or a technology or knowledge spill-over across diseases could all play a role in advancing knowledge integration. HIV, tuberculosis, and malaria being the main drivers of global integration, account for 52.1\% of total infectious disease funding from G20 countries in 2000-2017 while HIV alone accounts for 40.1\% of total funding with 42.1 billion US dollar \citep{head_allocation_2020}. The substantial resources specifically devoted to HIV research have led to collateral benefits to other disease areas \citep{schwetz_extended_2019}. Some examples include advancing antiviral drug development (on hepatitis C), improving vaccine research techniques (on ebola, zika, and influenza), enhancing understanding of immunology (on the role of $CD4+$ $T$ cells in fighting other infectious diseases and certain cancers) and advancing structural biology (on structure-based vaccine design that can be applied to other pathogens) \citep{schwetz_extended_2019}. Such unintentional but broad spin-offs might have enabled research on prominent diseases like HIV to bridge a wide range of bodies of knowledge and drive knowledge integration in the field of infectious disease research. Further disentanglement of different mechanisms behind integration might require identifying research fields and types of science of the publications on infectious diseases, and we leave this task to future research.


We also found that the number of publications on coronavirus has skyrocketed since COVID-19 emerged. The generated scientific knowledge of coronavirus research has been informing public health responses, treatments, and vaccine development \citep{micah_global_2023}. However, it has been argued that this surge reflects opportunism by both researchers and journals \citep{clark_how_2023}. For researchers, there has been a ``covidisation" of research to remain relevant and secure funding; for journals, there has been a loose ``gate-keeping" followed by fraudulent and poor quality research but eventually higher impact factors due to bulk citations \citep{clark_how_2023,glasziou_waste_2020}. Despite the surge in publications, the systemic impact of coronavirus on IDR was found to be fairly small. We suspected there might be a temporal delay for the systemic impact to catch up, but after validating the idea on the zika outbreak we found no evidence for such a delay. We encourage future research to further investigate this in depth.

\section{Conclusion}
This research provides a generalisable framework to examine the impact of funding in interdisciplinary knowledge creation. There is limited money available for global health research and development, especially from public and charitable funders. Thus, we must invest wisely. Effective research funding is vital in driving breakthroughs in science and technology. However, we caution that funded research output has witnessed a growing conservatism in the past decades, potentially slowing down scientific progress. We urge funding agencies and policymakers to better recognize and reward interdisciplinary contributions. Funding agencies and research institutions should prioritize mechanisms that encourage intelligent risk-taking.
\newpage
\section*{CRediT authorship contribution statement}
\textbf{Anbang Du:} Conceptualization, Software, Data curation, Resources, Formal Analysis, Methodology, Visualization, Writing – original draft,  Writing – review \& editing

\textbf{Michael Head:} Supervision, Data curation, Resources, Writing – review \& editing

\textbf{Markus Brede:} Conceptualization, Supervision, Methodology, Writing – review \& editing

\section*{Conflict of interest}
There is no conflict of interest for this manuscript.

\section*{Funding}
This research did not receive any specific grant from funding agencies in the public, commercial, or not-for-profit sectors.

\section*{Data availability}
Data and code are available on request.

\cleardoublepage

\appendix
\appendixpage
\counterwithin*{figure}{section}
\counterwithin*{table}{section}
\stepcounter{section}

\begin{xltabular}{\linewidth}{@{} l >{\hsize=2\hsize}X >{\hsize=.5\hsize}X >{\hsize=.5\hsize}X @{}}
        \hline\hline
        \endhead
        Disease    & Search Term   & A Records   & F Records     \\ 
        \hline\hline
         Coronavirus &   "COVID" OR "COVID-19" OR "Coronavirus" OR "Corona virus" OR "2019-nCoV" OR "SARS-CoV" OR "MERS-CoV" OR "Severe Acute Respiratory Syndrome" OR "Middle East Respiratory Syndrome"      &   284084   &  120371   \\ 
        \hline
        HIV       &   "HIV" OR ("AIDS" AND (immun* OR patient* OR epidem* OR pandemic*)) OR "Human immunodeficiency virus" OR "Acquired Immune Deficiency syndrome" OR "acquired immunodeficiency syndrome”   &  277242 & 143086   \\ 
        \hline
         Pneumonia          &   "Pneumonia" OR "pneumonias" OR ((lower respiratory tract infection*) OR (severe respiratory tract infection*)) &  106260 & 40569 \\ 
        \hline
          Tuberculosis &   "Tuberculosis"      &  102360 & 48078 \\ 

         \hline
         Influenza & ("flu" AND (pandemic* OR vaccin* OR shot* OR season*)) OR "influenza"&    74697 & 44141 \\
         \hline
         Hepatitis C & "Hepatitis C" OR ("hcv" AND (infect* OR virus* OR patient* OR hepatitis OR liver)) &    68295&28374 \\ 
         \hline
         Malaria & "Malaria" OR "Malarial" OR Plasmodium infect* &  65608 & 39227 \\           
         \hline
          Salmonella &  "Salmonella" & 61690 & 31563 \\ 
          \hline
          Hepatitis B  & "Hepatitis B" OR ("hbv" AND (infect* OR virus* OR patient* OR hepatitis OR liver)) &57374 & 26185 \\ 
          \hline
          Herpes&  ("HSV" AND (infect* OR vaccin* OR 1 OR 2 OR virus*)) OR "Herpes" OR "Shingles"    & 37844 & 17525\\ 
          \hline
          Urinary Tract Infection & "Urinary Tract Infect*" OR ("UTI" AND (E. coli OR antibiotic OR chlamydia OR Patient*)) & 30919 & 10037\\ 
          \hline
          Meningitis&"Meningitis"&27475&8692\\ 
          \hline
           Dengue&  "Dengue"  &   22639&13916 \\   
          \hline
            Chlamydia         &   "Chlamydia" OR "Chlamydiae" OR "Chlamydial"   &   17801 & 8137 \\ 
           \hline
           Leishmaniasis          &    "leishmaniasis"   &   17421&9651 \\ 

                     \hline
           Pertussis   &    "Pertussis" OR "whooping cough" & 15864 & 6943 \\   
                     \hline
           Measles          &    "Measles"     &    11464 & 4622\\ 
                     \hline
           Tetanus          &   "Tetanus"      &    10857&4462 \\ 
                     \hline
           Chagas          &   "chagas" OR "American trypanosomiasis"    &    10684    & 6126  \\ 
                     \hline
            Syphilis  & "syphilis" &10313& 3810\\   
                     \hline
            Varicella &"Varicella" OR "Chickenpox"&9049&3032\\ 
                     \hline
            Schistosomiasis & "Schistosomiasis" &8833&4685\\ 
                     \hline
            Zika         &"Zika"&   8182 & 6129\\ 
                     \hline
            Rabies        &  "Rabies"       &    7935 & 3540 \\ 
                     \hline
            Ebola        &   "Ebola" OR "Ebolavirus"      &7559&4565\\   
                     \hline
            Hepatitis A        &"Hepatitis A" OR ("hav" AND (infect* OR virus* OR patient* OR hepatitis OR liver))&7458 & 2367\\ 
                     \hline
            Diphtheria  & "Diphtheria" &7299&3536\\ 
                     \hline
             Leprosy        &"Leprosy"&6963&2262\\ 
                     \hline       
             Hepatitis E      &"Hepatitis E" OR ("hev" AND (infect* OR virus* OR patient* OR hepatitis OR liver))&5080&2747\\   
                     \hline
             Gonorrhoea&"N gonorrhoeae" infect* OR "Neisseria gonorrhoeae" Infect* OR "Gonorrhoea"&4932&2543\\ 
                     \hline
             Yellow fever&"yellow fever"&4245&2464\\ 
                     \hline
             filariasis &(filaria* AND (lymph* OR Elephantia* OR BANCROFTI* OR MALAYI OR Brugia*))&3723&1972\\ 
                     \hline
             trypanosomiasis        &"Sleeping Sickness" OR "African trypanosomiasis" OR "Trypanosoma brucei gambiense" OR "Trypanosoma brucei rhodesiense"&3571&2301\\ 
                     \hline
             Scabies        &"Scabies"&2126&615\\ 
                     \hline            
             Onchocerciasis &"Onchocerciasis"&1730&802\\ 
                     \hline
             Trichomoniasis &"Trichomoniasis"&1569&770\\ 
                     \hline\hline
    \caption{Search terms, total and funded number of records for each disease from 1995 to 2022. The data extraction was performed on 2024.03.29. }
    \label{D36}
\end{xltabular}

\begin{table}[htp]
    \centering
    \begin{tabular}{ll}
    \hline\hline
    Pair & False Positive Rate   \\
    \hline\hline
    HIV-TB & 2\%\\
    \hline
    Dengue-Zika & 1\%\\
    \hline
    Tetanus-Diphtheria & 0\%\\
    \hline
    HCV-HBV & 2\%\\
    \hline
    Clamydia-Gonorreahea & 3\%\\
    \hline\hline
    \end{tabular}
    \caption{False positive rate of knowledge integration in a paper's abstract considering five representative pairs of infectious diseases. The sample size for each pair is 100. The false positive rates are consistently less than 3\% and we found no bias towards particular disease pairs.}
    \label{false positive}
\end{table}

\begin{xltabular}{\linewidth}{@{}XX  @{}}
    \hline\hline
    \endhead
    Abbreviation & Disease  \\ 
    \hline\hline
Corona & Coronavirus \\ 
\hline
HIV & HIV \\ 
\hline
Pneum & Pneumonia \\ 
\hline
TB & Tuberculosis \\ 
\hline
Influenza & Influenza \\ 
\hline
HCV & Hepatitis C \\ 
\hline
Malaria & Malaria \\ \hline
Salm & Salmonella \\ \hline
HBV & Hepatitis B \\ \hline
Herpes & Herpes \\ \hline
UTI & Urinary Tract Infection \\ \hline
Mening & Meningitis \\ \hline
Dengue & Dengue \\ \hline
Chlamy & Chlamydia \\ \hline
Leishma & Leishmaniasis \\ \hline
Pert & Pertussis \\ \hline
Measles & Measles \\ \hline
Teta & Tetanus \\ \hline
Chagas & Chagas \\ \hline
Syphilis & Syphilis \\ \hline
Vari & Varicella \\ \hline
Schisto & Schistosomiasis \\ \hline
Zika & Zika \\ \hline
Rabies & Rabies \\ \hline
Ebola & Ebola \\ \hline
HAV & Hepatitis A \\ \hline
Diph & Diphtheria \\ \hline
Leprosy & Leprosy \\ \hline
HEV & Hepatitis E \\ \hline
Gono & Gonorrhoea \\ \hline
YF & Yellow Fever \\ \hline
Fila & Filariasis \\ \hline
Trypano & Trypanosomiasis \\ \hline
Scabies & Scabies \\ \hline
Onchocer & Onchocerciasis \\ \hline
Trichomo & Trichomoniasis \\ \hline
\hline
    \caption{Disease names and abbreviations used in this study.}
    \label{abbr}
\end{xltabular}

\begin{table}[htp]
\centering
\begin{tabular}{lllllll}
\hline\hline
Year & WOS F\% & 36 F\% & All WOS & F WOS & All 36 & F 36 \\
\hline\hline
1995 & 0.06 & 0.18 & 798461 & 43966 & 18818 & 3372 \\
\hline
1996 & 0.05 & 0.17 & 886477 & 47519 & 22566 & 3801 \\
\hline
1997 & 0.05 & 0.17 & 898426 & 47278 & 22952 & 3853 \\
\hline
1998 & 0.05 & 0.17 & 923073 & 47735 & 23595 & 4097 \\
\hline
1999 & 0.05 & 0.17 & 897754 & 48810 & 24177 & 4161 \\
\hline
2000 & 0.05 & 0.17 & 939156 & 47548 & 24608 & 4063 \\
\hline
2001 & 0.05 & 0.16 & 933384 & 48926 & 24073 & 3922 \\
\hline
2002 & 0.05 & 0.16 & 951628 & 50729 & 24596 & 3910 \\
\hline
2003 & 0.05 & 0.16 & 1008291 & 54242 & 25677 & 4156 \\
\hline
2004 & 0.06 & 0.17 & 1050680 & 59610 & 27334 & 4727 \\
\hline
2005 & 0.06 & 0.19 & 1179694 & 75764 & 29187 & 5622 \\
\hline
2006 & 0.11 & 0.21 & 1251501 & 140190 & 31235 & 6520 \\
\hline
2007 & 0.12 & 0.21 & 1371632 & 158111 & 33888 & 7168 \\
\hline
2008 & 0.22 & 0.34 & 1527949 & 328555 & 36557 & 12463 \\
\hline
2009 & 0.39 & 0.53 & 1638275 & 635089 & 39573 & 21012 \\
\hline
2010 & 0.43 & 0.58 & 1688059 & 730003 & 43165 & 25125 \\
\hline
2011 & 0.46 & 0.61 & 1796994 & 820457 & 46694 & 28267 \\
\hline
2012 & 0.47 & 0.62 & 1902726 & 893601 & 48418 & 30011 \\
\hline
2013 & 0.48 & 0.64 & 2000048 & 960986 & 50688 & 32378 \\
\hline
2014 & 0.48 & 0.64 & 2108895 & 1007809 & 51995 & 33419 \\
\hline
2015 & 0.48 & 0.65 & 2191395 & 1059818 & 53713 & 35027 \\
\hline
2016 & 0.51 & 0.65 & 2295920 & 1165592 & 55004 & 35863 \\
\hline
2017 & 0.57 & 0.67 & 2381988 & 1354334 & 56551 & 37834 \\
\hline
2018 & 0.58 & 0.67 & 2442581 & 1416402 & 56792 & 38025 \\
\hline
2019 & 0.59 & 0.66 & 2644412 & 1564103 & 59708 & 39574 \\
\hline
2020 & 0.60 & 0.52 & 2795066 & 1672916 & 110972 & 57726 \\
\hline
2021 & 0.62 & 0.51 & 3023128 & 1871583 & 181063 & 91837 \\
\hline
2022 & 0.62 & 0.51 & 3078441 & 1911288 & 188602 & 95997 \\
\hline
\end{tabular}
\caption{The number and proportion of funded publications through time in the WOS Core Collection. \textbf{All WOS}: the total number of WoS publications (filtered by SQ1). \textbf{F WOS}: all funded WoS publications (filtered by SQ1 and SQ2). \textbf{WOS F\%}: the proportion of funded WoS publications. \textbf{All 36}: the total number of WoS publications on the 36 diseases (filtered by SQ1 and all 36 diseases' search terms joined by 'OR' in the Topic field). \textbf{F 36}: the total number of WoS publications on the 36 diseases (filtered by SQ1 and SQ2 and all 36 diseases' search terms joined by 'OR' in the Topic field). \textbf{36 F\%}: the proportion of funded WoS publications on the 36 diseases. Data extraction was done on 10/04/2024. The drop in 36F\% starting the year 2020 was due to the mass increase in coronavirus publication coupled with a low rate of identified funded research as shown in \tref{Covid_perc}.}
\label{Fperc}
\end{table}

\begin{table}[htp]
\centering
\begin{tabular}{cccccc}
\hline\hline
Year & Corona & Corona F & Corona 36\% & Corona F\% & 36 F\% \\ \hline\hline
1995 & 116 & 32 & 0.01 & 0.28 & 0.18 \\ \hline
1996 & 96 & 24 & 0.00 & 0.25 & 0.17 \\ \hline
1997 & 112 & 27 & 0.00 & 0.24 & 0.17 \\ \hline
1998 & 138 & 28 & 0.01 & 0.20 & 0.17 \\ \hline
1999 & 101 & 23 & 0.00 & 0.23 & 0.17 \\ \hline
2000 & 95 & 20 & 0.00 & 0.21 & 0.17 \\ \hline
2001 & 122 & 29 & 0.01 & 0.24 & 0.16 \\ \hline
2002 & 82 & 15 & 0.00 & 0.18 & 0.16 \\ \hline
2003 & 325 & 33 & 0.01 & 0.10 & 0.16 \\ \hline
2004 & 775 & 59 & 0.03 & 0.08 & 0.17 \\ \hline
2005 & 711 & 91 & 0.02 & 0.13 & 0.19 \\ \hline
2006 & 615 & 122 & 0.02 & 0.20 & 0.21 \\ \hline
2007 & 454 & 101 & 0.01 & 0.22 & 0.21 \\ \hline
2008 & 447 & 166 & 0.01 & 0.37 & 0.34 \\ \hline
2009 & 393 & 259 & 0.01 & 0.66 & 0.53 \\ \hline
2010 & 372 & 247 & 0.01 & 0.66 & 0.58 \\ \hline
2011 & 310 & 225 & 0.01 & 0.73 & 0.61 \\ \hline
2012 & 308 & 231 & 0.01 & 0.75 & 0.62 \\ \hline
2013 & 380 & 288 & 0.01 & 0.76 & 0.64 \\ \hline
2014 & 474 & 349 & 0.01 & 0.74 & 0.64 \\ \hline
2015 & 497 & 330 & 0.01 & 0.66 & 0.65 \\ \hline
2016 & 541 & 366 & 0.01 & 0.68 & 0.65 \\ \hline
2017 & 553 & 410 & 0.01 & 0.74 & 0.67 \\ \hline
2018 & 507 & 391 & 0.01 & 0.77 & 0.67 \\ \hline
2019 & 583 & 441 & 0.01 & 0.76 & 0.66 \\ \hline
2020 & 48666 & 16710 & 0.44 & 0.34 & 0.52 \\ \hline
2021 & 116651 & 49424 & 0.64 & 0.42 & 0.51 \\ \hline
2022 & 127050 & 57309 & 0.67 & 0.45 & 0.51 \\ \hline
\end{tabular}
\caption{The number and proportion of funded Coronavirus publications through time in the WoS. \textbf{Corona}: total number of Coronavirus-related publications (filtered by SQ1). \textbf{Corona F}: total number of Coronavirus-related funded publications (filtered by SQ1 and SQ2). \textbf{Corona 36\%}: proportion of Coronavirus-related publications within all WoS publications on the 36 diseases. \textbf{Corona F\%}: proportion of funded Coronavirus-related publications out of all Coronavirus-related publications. \textbf{36 F\%}: the proportion of funded WoS publications on the 36 diseases. Data extraction was done on 16/04/2024.}
\label{Covid_perc}
\end{table}

\begin{figure}[htp]
    \centering    
    \includegraphics[width=\textwidth]{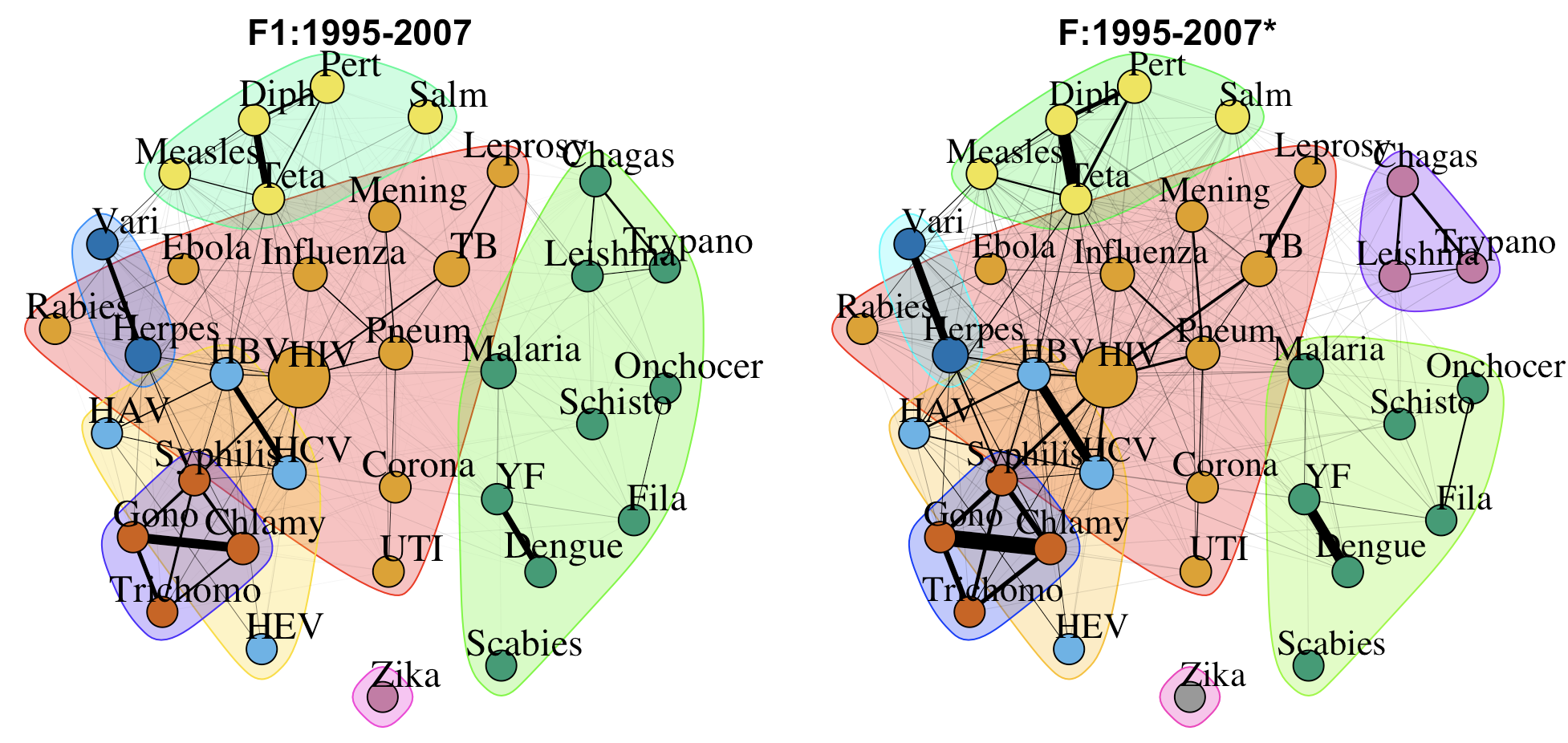}
    \caption{Comparison of the average funded network 1995-2007 including or not including(*) the years 1997 and 2004. There is no major difference in the network structure after including these two years, except that certain links get slightly weakened due to the noise introduced.}
    \label{Fig7}
\end{figure}

\begin{figure}[htp]
    \centering    
    \includegraphics[width=\textwidth]{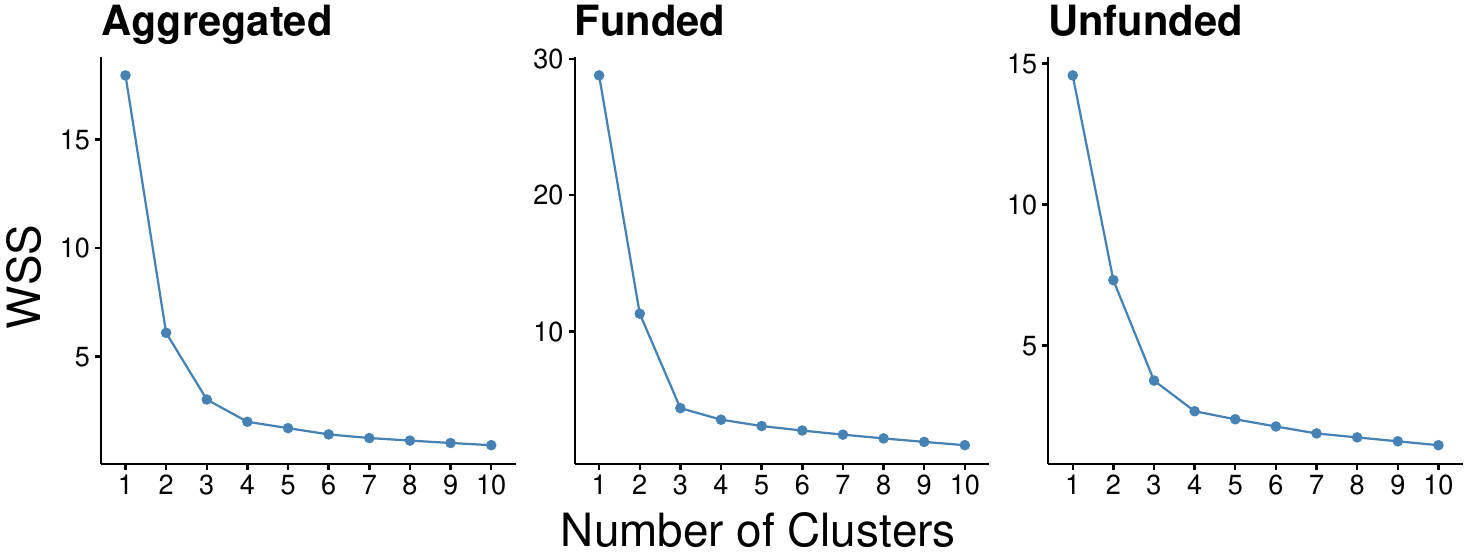}
    \caption{Elbow method of identifying the optimal number of clusters. The within-cluster sum of squares (WSS) for different cluster numbers is measured, where the WSS of a cluster is its sum of squares to the centroid, i.e., WSS $= \sum_{i=1}^{k}\sum_{x\in C_i}|x-c_i|^2$, with $C_i$ the set of years belonging to cluster $i$'s and $c_i$ the clusters centroid. The funded elbow plot is created by removing the years 2004 and 1997. We observe that choosing three clusters would be a suitable choice for funded, unfunded, and all research.}
    \label{Fig8}
\end{figure}

\begin{figure}[htp]
    \centering
    \includegraphics[width=\textwidth]{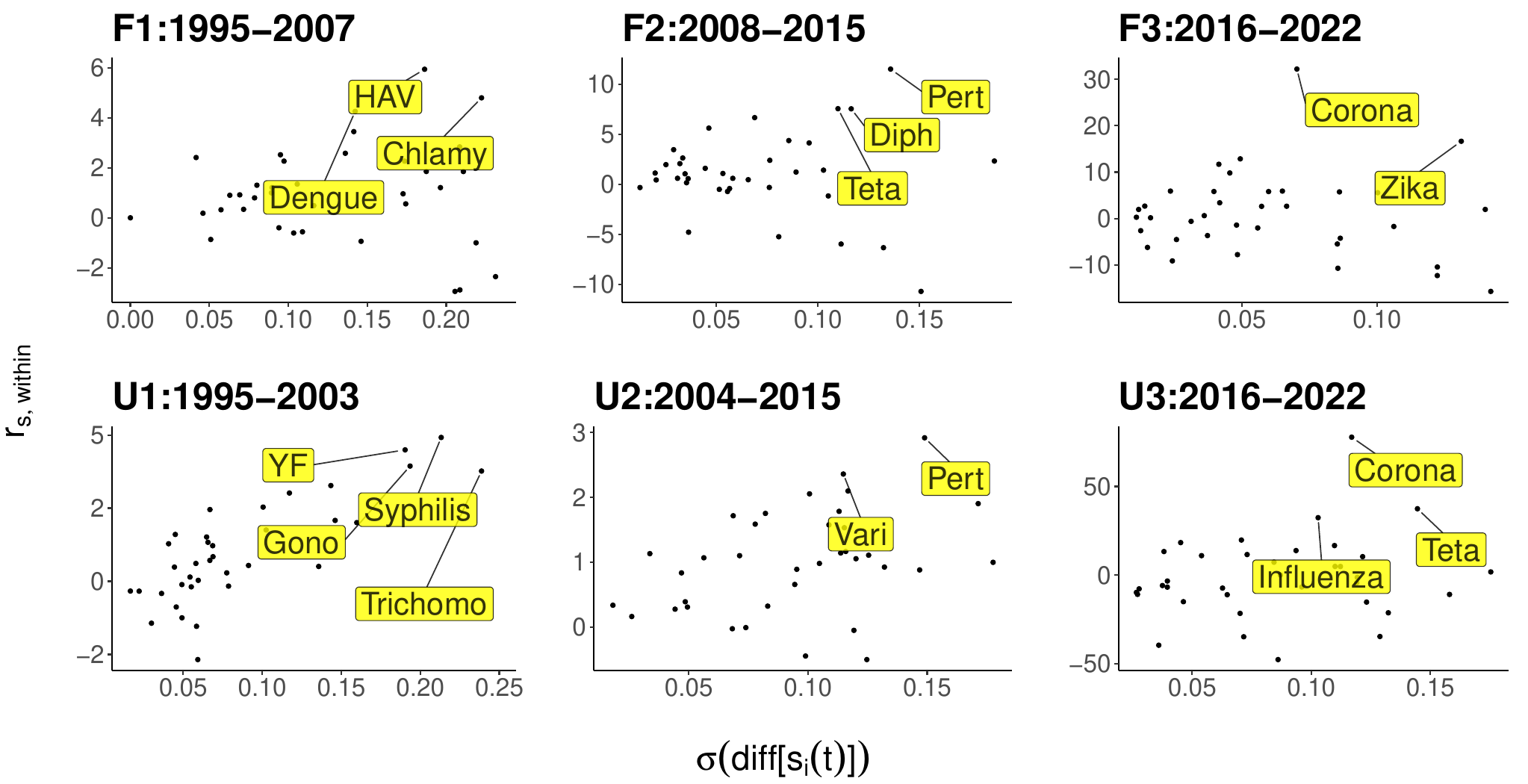}
    \caption{Analysis of contributions of individual diseases to system change within funded regimes (F1:1995-2007, F2:2008-2015, F3:2016-2022) and unfunded regimes (U1:1995-2003, U2:2004-2015, U3:2016-2022) in terms of relative change in node strength $r_{s,within}$ plotted against volatility $\sigma(diff[s_i(t)])$. The diseases with the highest $r_{s,within}$ were highlighted in each regime. Hepatitis A (HAV), chlamydia and dengue were highlighted in F1 while yellow fever, gonorrhoea, syphilis, and trichomoniasis in U1; DTP in F2 while varicella and pertussis in U2; coronavirus and zika in F3 while coronavirus, tetanus and influenza in U3.}
    \label{Fig3}
\end{figure}

\begin{figure}[htp]
    \centering
    \includegraphics[width=\textwidth]{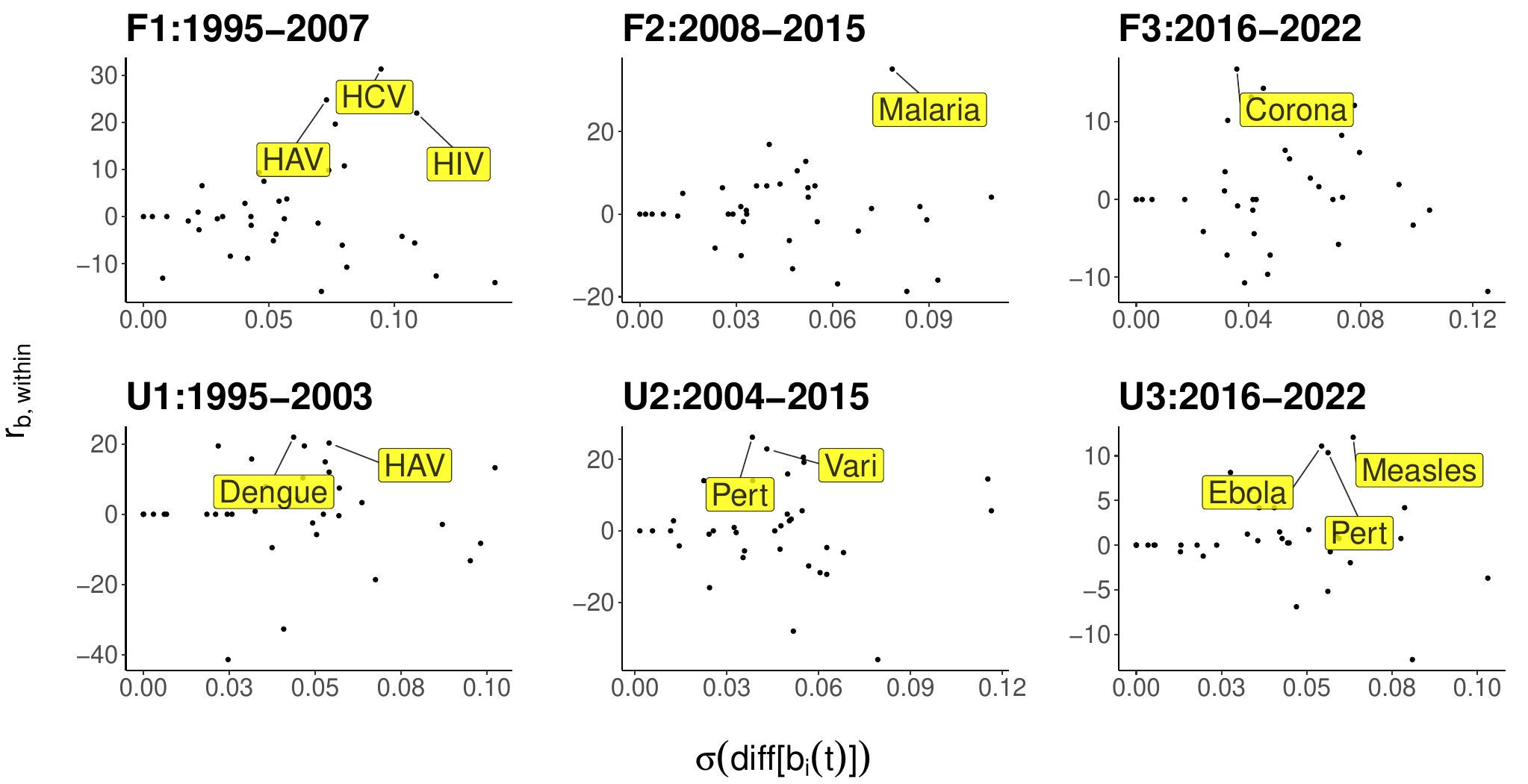}
    \caption{Analysis of contributions of individual diseases to system change within F regimes (F1:1995-2007, F2:2008-2015, F3:2016-2022) and U regimes (U1:1995-2003, U2:2004-2015, U3:2016-2022) in terms of relative change in betweenness $r_{b, within}$ versus volatility $\sigma(diff[b_i(t)])$. The diseases with the highest $r_{b,within}$ were highlighted in each regime. HCV, HAV and HIV were highlighted in F1 while dengue and HAV in U1; malaria in F2 while varicella and pertussis in U2; coronavirus in F3 while measles, ebola and pertussis in U3.}
    \label{Fig5}
\end{figure}

\begin{figure}[htp]
    \centering    
    \includegraphics[width=\textwidth]{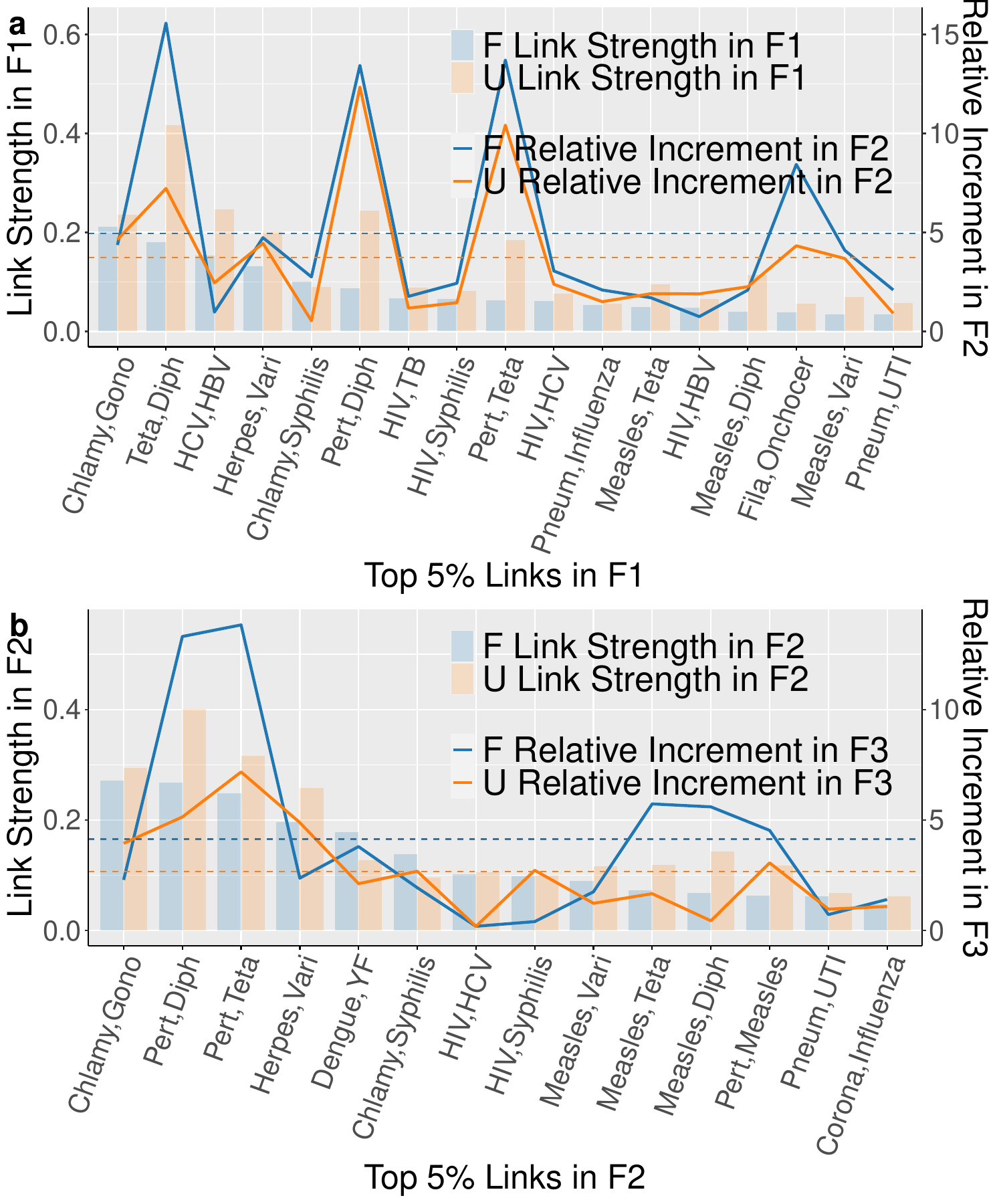}
    \caption{Comparison of relative increments of the top 5\% strongest links. Bar height represents the average link strength of the corresponding disease pair in the first period, and the solid line represents the pair's increment in strength to the second period, relative to the average change in link strength. The dashed lines give average relative increments for all pairs included in the figure for funded (blue) and unfunded (orange) research. (a) Change from F1 (1995-2008) to F2 (2009-2015) with average relative increments of 4.95 for funded and 3.74 for unfunded research. (b) Change from F2 (2009-2015) to F3 (2016-2022), with average relative increments of 4.13 for funded and 2.67 for unfunded research.}
    \label{Fig9}
\end{figure}

\begin{figure}[htp]
    \centering    
    \includegraphics[width=\textwidth]{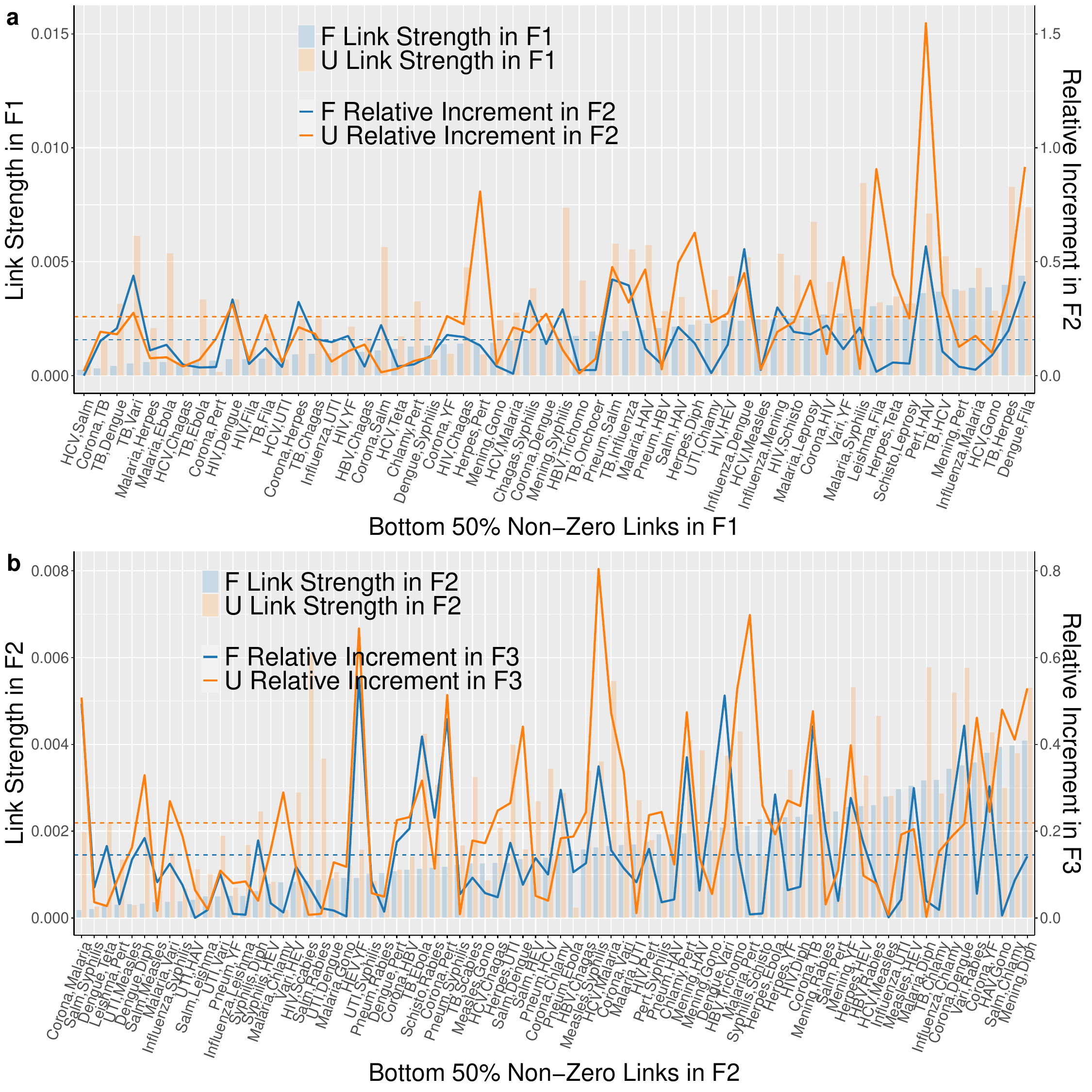}
    \caption{Comparison of relative increments of the bottom 50\% weakest (non-zero) links. Bar height represents the average link strength of the corresponding disease pair in the first period, and the solid line represents the pair's increment in strength to the second period, relative to the average change in link strength. The dashed lines give average relative increments for all pairs included in the figure for funded (blue) and unfunded (orange) research. (a) Change from F1 (1995-2008) to F2 (2009-2015), with average relative increments of 0.16 for funded and 0.26 for unfunded research. (b) Change from F2 (2009-2015) to F3 (2016-2022), with average relative increments of 0.15 for funded and 0.22 for unfunded research.}
    \label{Fig10}
\end{figure}
\clearpage





\end{document}